\documentclass{JAIS}
\usepackage{hyperref}

\journal{JAIS-ID}
\vol{2022}

\received{xx January 2018}
\published{xx March 2018}

\def\be{\begin{equation}}
\def\ee{\end{equation}}
\def\bea{\begin{eqnarray}}
\def\eea{\end{eqnarray}}

\usepackage{caption}
\usepackage{lineno}

\articletype{Technical Report}

\begin{document}

\title{Using Micromegas detectors for direct dark matter searches: challenges and perspectives}

\author{K. Altenm\"{u}ller\auno{1}, I. Antolín\auno{1},  D. Calvet\auno{2}, F. R. Candón \auno{1}, J. Castel\auno{1}, S. Cebrián\auno{1}, C. Cogollos\auno{3,4}, T. Dafni\auno{1}, D. Díez Ibáñez\auno{1}, E. Ferrer-Ribas\auno{2} J. Galán\auno{1}, J.A. García\auno{1}, H. Gómez\auno{2}, Yikun Gu\auno{1}, A. Ezquerro\auno{1}, I.G Irastorza\auno{1}, G. Luzón\auno{1}, C. Margalejo\auno{1}, H. Mirallas\auno{1},  L. Obis\auno{1}, A. Ortiz de Solórzano\auno{1}, T. Papaevangelou\auno{2}, O. Pérez\auno{1}, E. Picatoste\auno{2},J. Porrón\auno{1},  M. J. Puyuelo \auno{1}, A. Quintana\auno{1,2}, E. Ruiz-Chóliz\auno{5},  J. Ruz\auno{1} and J. Vogel\auno{1}}
\address{$^1$Centro de Astropartículas y Física de Altas Energías (CAPA), Universidad de Zaragoza, Spain.\\
$^2$Université Paris-Saclay, CEA, IRFU, 91191 Gif-Sur-Yvette, France. \\
$^3$ICCUB Instituto de Ciencias del Cosmos - Universidad de Barcelona, Spain. \\
$^4$Max-Planck-Institut f\"ur Physik (Werner-Heisenberg-Institut), Germany. \\
$^5$Institut f\"ur Physik, Johannes Gutenberg Universit\"at Mainz, Germany.

\vspace{0.2cm}
Corresponding author: G. Luzón\\
Email address: luzon@unizar.es}

\begin{abstract}
Gas time projection chambers (TPCs) with Micromegas pixelated readouts are being used in dark matter searches and other rare event searches, due to their potential in terms of low background levels, energy and spatial resolution, gain, and operational stability. Moreover, these detectors can provide precious features, such as topological information, allowing for event directionality and powerful signal-background discrimination. The Micromegas technology of the microbulk type is particularly suited to low-background applications and is being exploited by detectors for CAST and IAXO (solar axions) and TREX-DM (low-mass WIMPs) experiments.
Challenges for the future include reducing intrinsic background levels, reaching lower energy detection levels, and technical issues such as robustness of detector, new design choices, novel gas mixtures and operation points, scaling up to larger detector sizes, handling large readout granularity, etc. We report on the status and prospects of the development ongoing in the context of IAXO and TREX-DM experiments, pointing to promising perspectives for the use of Micromegas detectors in direct dark matter searches.
\end{abstract}

\maketitle


\section{Introduction}

The nature of Dark matter (DM) is one of the most intriguing and elusive phenomena in modern physics. Despite its overwhelming abundance in the universe, accounting for about 27\% of its energy density, DM's nature and properties remain unknown. Various theoretical models propose that DM consists of weakly interacting massive particles (WIMPs) that could be produced in the early stages of the Big Bang. Direct detection experiments for WIMPs aim to observe the scattering of WIMPs from our galactic halo as they interact with nuclei in a low-background underground detector, measuring the recoil energy. Mainstream WIMP experiments rely on the properties of WIMPs suggested by the so-called ``WIMP miracle'', which refers to the non-trivial coincidence that particles with masses around O(100)\,GeV and annihilation cross-sections at the electroweak scale produce relic densities roughly corresponding to the observed dark matter density.  However, despite enormous progress in experimental sensitivity, so far there is no sign of WIMPs in underground experiments. This fact, together with the absence of clear clues from indirect DM searches or colliders, leads researchers to shift focus to other approaches or candidates. One interesting option is that of WIMPs with masses substantially below 10\,GeV, which could be motivated by particular phenomenological scenarios, notably the asymmetric DM paradigm \cite{Blennow_2012,ZUREK201491}. Their nuclear recoils would lie below conventional energy thresholds and would have remained undetected so far. Moreover, altogether different DM candidates at the low mass frontier, like axions or axion-like particles, are emerging as very attractive alternatives. Axions are very well motivated by theory, as they solve the strong-CP problem of the Standard Model \cite{PQ_Axion}. Despite being very light particles (masses below eV), axions would be produced non-thermally and non-relativistically at early times, and therefore are excellent DM candidates. Although direct detection of DM axions employs techniques very different than those of the WIMP detectors, the fact that axions are produced in the Sun offers other detection opportunities without parallel in the WIMP sector. Experiments looking for solar axions, known as axion helioscopes, trigger their conversion into X-rays in strong laboratory magnetic fields. The detection of the resulting X-rays shares similar experimental requirements to low-mass WIMP detectors, given that both need very low background in the keV region. The material discussed below is framed in both low-mass WIMP and solar axion searches.

Most of the conventional direct detection WIMP experiments, targeting relatively heavy (i.e. ``WIMP miracle'' inspired) WIMPs, are based on their capability to accumulate a large quantity (several tons) of target mass and exposure, and their ability to discriminate nuclear recoil signals from other backgrounds, thus reaching an extraordinarily low background level. At the moment, the most successful technique in doing so are noble liquid detectors (typically in double phase), e.g.~\cite{xenon1t,darkside2017}, having reached the highest sensitivity to O(100)\,GeV WIMPs as DM candidates. However, if the WIMP mass is sufficiently low, the nuclear recoil produced in the detector will be below the discrimination threshold  (and eventually below the hardware threshold of the detector). A new category of experiments, targeting low-mass WIMPs of few GeV and below, have emerged during the last years. These experiments prioritize low energy threshold over large exposure. Without discriminating capabilities, they face higher backgrounds and the typical target masses are still at the kg scale. Moreover, maximizing the radiopurity of materials and understanding the systematics of the detection right at the threshold are crucial challenges in this quest. 

The experimental race towards the low-mass WIMPs includes, among others, small size bolometer arrays (as  CRESST-III \cite{CRESST2019}, CDMSLite \cite{CDMSLite2018} or Edelweiss \cite{Edelweiss2019}), point-contact germanium semiconductor detectors (as  CDEX \cite{CDEX2019}),  or highly pixelized Si detectors (as DAMIC \cite{DAMIC2020}). In this context, as will be argued below, gaseous Time Projection Chambers (TPC) with Micro-Pattern Gas Detectors (MPGD) readout planes also constitute very promising options, as they combine the low threshold allowed by the charge amplification in gas, together with good scalability prospects and a rich event information thanks to the patterned readout. In fact, gaseous TPCs have long been explored in the context of WIMP searches, but mostly focused on their potential as directional DM detectors, in which the WIMP-induced nuclear recoil track is imaged and its direction determined. In the case of a positive detection, the directional signal of a WIMP is considered the most unmistakable signature of a DM origin. But directional sensitivity, although possibly only realistic in gas TPCs, is still very challenging. In order to adequately image few keV recoils, radical TPC implementations are needed, like e.g. the operation at very low pressure (of O(100)\,Torr), very high granularity readouts or negative-ion drift. In this context, MPGDs, and Micromesh Gas Structures (Micromegas) in particular, are being explored by projects like DRIFT~\cite{drift}, MIMAC~\cite{mimac} or CYGNO~\cite{cygno}. These efforts have resulted in successful prototypes. However, the realization of a full-scale experiment with competitive sensitivity remains challenging, but certainly something to be attempted if, in the future, a positive signal is seen in a non-directional experiment. Directional-specific challenges will not be reviewed in this paper, in which we focus on the more conventional application of Micromegas readout planes in (non-directional) low-mass WIMP detectors, in which somewhat high pressure is preferred as a way to increase exposure.

Experiments like  NEWS-G \cite{newsg} and  TREX-DM \cite{trexdm2016},  the latter equipped with Micromegas, are both attempting to search for low-mass WIMPs using high-pressure gaseous targets. The charge amplification occurring at the TPC anode offers a natural strategy for reaching very low energy thresholds, which are decoupled from the detector size (contrary to many other detection technologies in which the detector size determines the threshold via the capacitive noise), and thus also with good scalability prospects. The traditional complexity associated with conventional TPCs (mechanical and electronics-wise) is however not a good match for the high radiopurity specifications of DM experiments. NEWS-G solves this issue by going to an innovative spherical geometry in which the anode ``plane'' is reduced to a small spherical electrode placed at the center of the detector. The utmost simplicity of the detector comes at the expense of less event information. In its turn, TREX-DM relies on the advances achieved in modern MPGDs, in particular Micromegas, in terms of robustness and simplicity of construction, while keeping the conceptual complexity of a highly pixelated readout. The challenges in developing and implementing Micromegas in experiments like TREX-DM is the main theme of this paper. Central to these efforts is the possibility of building MPGDs out of very radiopure materials, like is the case of microbulk Micromegas planes.

As already mentioned, the detection needs in axion helioscopes share similar features with the low-mass WIMP experiments, as both pursuits confront the challenge of extremely low rates and low energy ($\sim$keV) levels for the expected signal events. Indeed, the technology developed for TREX-DM readouts was itself based on an early low-background application in the CAST axion helioscope \cite{García_2013}. These detectors are now being further improved, as baseline option for the focal point detectors of the future BabyIAXO and IAXO axion helioscopes \cite{abeln2021conceptual}, as will be reviewed later on.

In the first section we will briefly review how these detectors have been used in experiments like CAST and how they are being prepared for IAXO, both devoted to axion searches; and in TREX-DM, looking for low-mass WIMP signals. Then, we will report on progresses: in lowering the energy threshold (section \ref{sec:lowThreshold}), lowering backgrounds (section \ref{sec:background}), and other technical issues (section \ref{sec:others}), along with an analysis on how these improvements have had an impact on TREX-DM sensitivity (section \ref{sec:TREXDMsensitivity}). We will finish with a summary on future perspectives and their impact on sensitivity (section \ref{sec:TREXDMsensitivity}) and some conclusions (section \ref{sec:conclusions}).

\section{Micromegas in direct DM searches: axions (CAST/IAXO) and WiMPs (TREX-DM)}\label{sec:DM-searches}

The last decades have witnessed significant advances of the development of gaseous Time Projection Chambers (TPCs) based on Micromegas technology. A review on  current status and future development of  Micromegas detectors for physics and applications with references can be found here \cite{MMs}. Here we will focus on direct DM searches.

Many of the ideas and developments presented in this section have their origin in the T-REX project which embarqued in an intensive R\&D on low background applications of gaseous TPCs with the aim of studying the applicability of Micromegas readouts TPCs to rare event searches (WIMP searches, axions and double beta decay)~\cite{Irastorza:2015dcb,Irastorza:2015geo}. Here we will focus on axion searches, summarizing the role of Micromegas in CAST and prospects for IAXO/BabyIAXO, and on the potential of these detectors for low mass WIMPs in the TREX-DM experiment.   

Axion helioscopes utilize powerful magnets aligned towards the Sun to convert solar axions into X-rays. The energy distribution of these photons follows that of the solar axions, which typically lies at the 1-10\,keV range, peaking around 4\,keV. The CERN Axion Solar Telescope (CAST), the most powerful axion helioscope to date, was operational at CERN for almost two decades. It used a 9\,Tesla prototype magnet from the Large Hadron Collider (LHC), with a length of  9.3\,m, 2\,apertures of 15\,cm$^2$ and the ability to track the Sun for about 3\,hours daily through elevation and azimuth drives. CAST has not detected any solar axion signals, setting upper limits on photon-axion coupling \cite{nature-2017,CAST-He3,CAST-He4}.

\begin{figure}[htbp]
\centering
\includegraphics[width=0.5\textwidth]{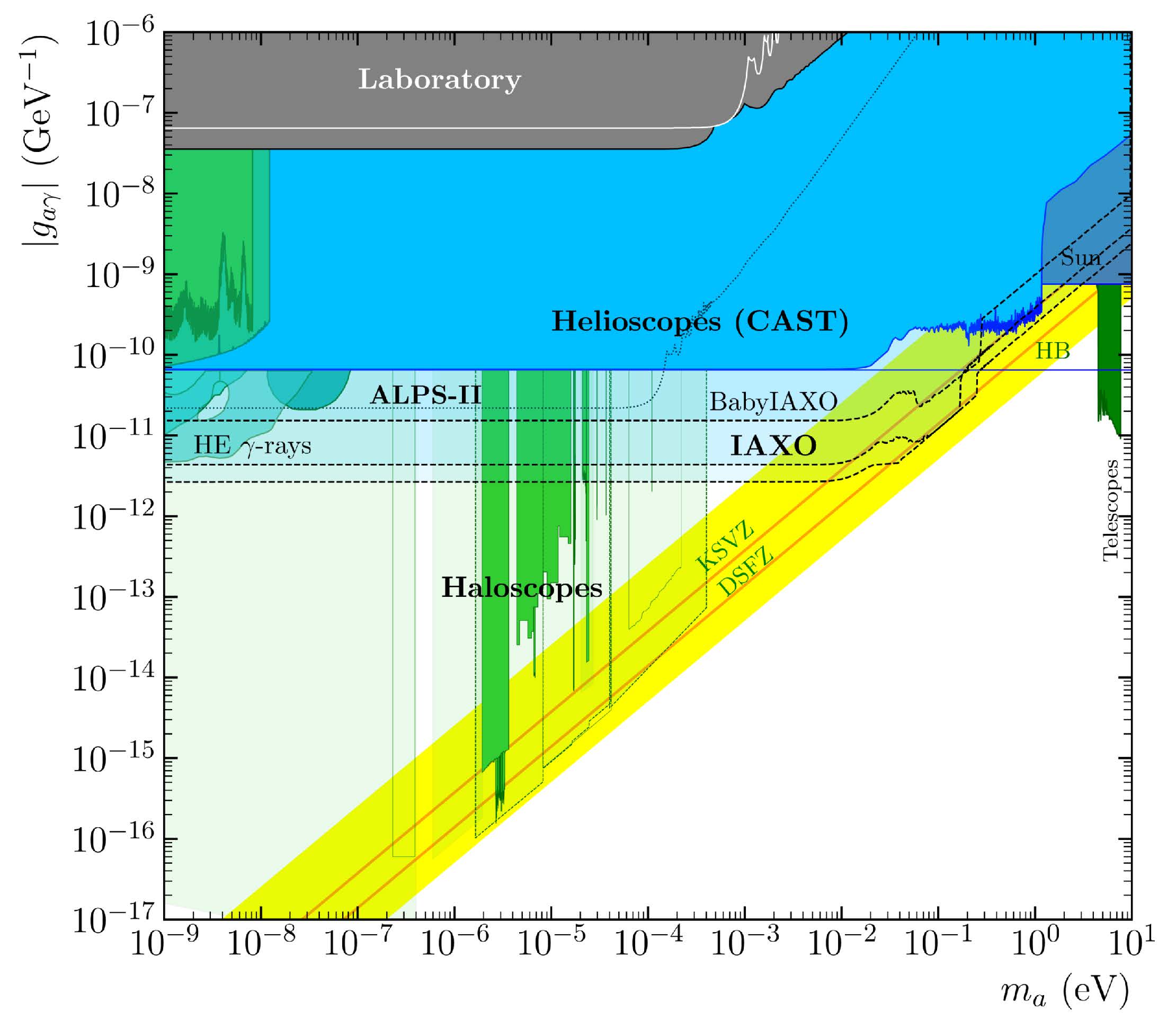}
\caption{Sensitivity plot of IAXO and BabyIAXO, in the $g_a$-$m_a$ parameter space, showing the  QCD axion (yellow) band, the CAST bound, and other current (solid) and future (dashed)
experimental and observational limits. See~\cite{abeln2021conceptual} for more details on the plot.\label{fig:Axion_Exclusion_Plot}}
\end{figure}

The International Axion Observatory (IAXO) \cite{Armengaud_2014} is the next generation axion helioscope designed to detect solar axions and axion-like particles (ALPs). This project is focused on constructing a sizeable 20\,m-long toroidal magnet comprising eight superconducting coils, specifically optimized for axion research. Within this magnet, eight 60\,cm-diameter bores equipped with X-ray optics will focus signal photons into 0.2\,cm$^2$ spots, captured by ultra-low background X-ray detectors. Detection technologies like microbulk Micromegas, similar to those in CAST, are being considered. The magnet, optics and detectors are mounted on a drive system to track the sun for approximately 12 hours a day, on average.

BabyIAXO \cite{abeln2021conceptual} is intended as a test platform for the components of IAXO (like the magnet, optics, and detectors) at a scale relevant to the final system. Additionally, it functions as a fully operational helioscope capable of exploring important physics on its own and possibly making new discoveries. The BabyIAXO magnet will have two 10\,m-long, 70\,cm-diameter bores, accommodating detection lines (optics and detectors) similar in size to those planned for IAXO. Figure \ref{fig:Axion_Exclusion_Plot} compares CAST and IAXO/BabyIAXO sensitivities~\cite{Armengaud_2019}.

\begin{figure}[htbp]
\centering
\includegraphics[width=0.8\textwidth]{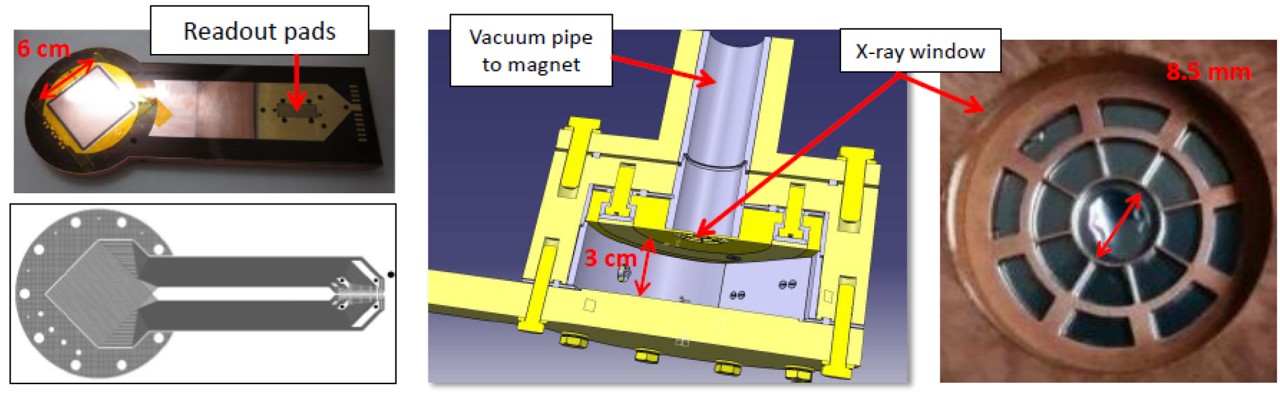}
\caption{Drawing and pictures of the CAST microbulk Micromegas \cite{Garza_2015}. Detector sketch showing the main components of the detection chamber (center). The 36\,cm$^2$ active area covering the 14.55\,cm$^2$ projection of the magnet’s bore, whose pixels are read in X and Y columns (120$\times$120 strips) using DAQ electronics based on the AFTER chip (left). The
X-ray window: a gas-tight, 4--5\,$\mu$m aluminized mylar\textsuperscript{\textregistered} foil glued on a spider-web patterned copper
strong-back that withstands the differential pressure to the magnet vacuum system (right).\label{fig:CAST MM}}
\end{figure}

Maximizing the figure of merit of axion helioscopes~\cite{NGAH} requires 
low-background X-ray detection methods and potential X-ray optics for better signal-to-noise ratios.
CAST used extensively Micromegas detectors to detect low-energy X-rays since the first stages of the experiment \cite{Abbon_2007, SAune_2013, SAune_2014,Aznar_2015}. While evolving over time, Micromegas detectors have maintained their basic design: a small TPC with a Micromegas readout (figure \ref{fig:CAST MM}). X-rays enter the detector via a gas-tight window, that is also the cathode of the TPC, and initiate ionization in the conversion volume, which has been designed to stop the signal photons, while minimizing background: typically of a length of 3\,cm height and filled with  argon at 1.4\,bar in addition to a small quantity of
quencher (e.g. 2\%\,isobutane).
The ionization cloud drifts to the Micromegas readout, where signal amplification occurs. These detectors have a finely pixelized (typical pitch of $\sim$0.5\,mm) readout plane, providing detailed 3D information about the ionization cloud. The small total active area of 36\,cm$^2$ easily contains the potential signal, concentrated on a few mm$^2$ thanks to the use of X-ray optics (see fig. 15 of \cite{Armengaud_2019}).

The success of these detectors in CAST is attributed to several factors:
\begin{itemize}
\item Micromegas readouts excel in radiopurity through the microbulk technique, using only highly pure materials like kapton\textsuperscript{\textregistered} and copper. They allow primary signals to be extracted from the detector to a distant and shielded point for front-end electronics, avoiding potential radioactivity sources like soldering or connectors. All components, including the chamber's body, X-ray window, and connectors, undergo rigorous screening and are made from radiopure materials (see section \ref{sec:background}).

\item These detectors offer detailed imaging of ionization signatures in gas, enabling the identification of signal-like events. The patterned anode and digitized temporal wave-forms of the pulses induced in each of the patterned strips form the basis for developing advanced algorithms that distinguish signal X-ray events from other background events.

\item To minimize external background sources, these detectors can be shielded passively and actively. While borrowing concepts from underground experiments, considerations must be made for specific constraints like the magnet's space and weight, in the presence of cosmic rays due to operation at the surface, and the inherent sensitivity and rejection capability of the Micromegas detectors.
\end{itemize}

The Micromegas detectors proposed for BabyIAXO are based on previous CAST detectors, with a focus on enhancing radiopurity. They have been built with 18\,mm thick radiopure copper walls and PTFE gaskets. A kapton\textsuperscript{\textregistered} field shaper is included to improve drift field uniformity and decrease border effects. One prototype, IAXO-D0, is already in operation at the University of Zaragoza, housed in a shielded setup with a 20\,cm thick lead castle. It is equipped with new data-acquisition (DAQ) electronics based on the AGET chip~\cite{AGET} as substitute of the AFTER (ASIC For TPC Electronics Readout)-based front-end card (FEC)~\cite{Abbon_2007}. The AGET has an autotrigger function, which allows to create the trigger from individual strips without the need of the mesh signal serving as external trigger: the mesh presents a much higher capacitance than each of the anode strips, thus, using this characteristic of the AGET chip, the energy threshold was improved. A closed recirculation gas system  is used in view of the use of Xe-based gas mixtures. The main improvement with respect to the CAST Micromegas detector is the thicker lead shielding and a 4$\pi$ veto shielding specially designed to tag cosmic-induced events. A second prototype has been installed underground at the Laboratorio Subterráneo de Canfranc (LSC), to study its performance in the absence of cosmic rays.

Regarding WIMP searches, gaseous TPCs offer versatility in target elements, robust tracking capabilities, and achieve remarkably low thresholds due to gas amplification. Moreover, recent advancements in electronics and novel radiopure  Micromegas, improve the scalability prospects (section \ref{sec:others}) and minimize the background (section \ref{sec:TREXDMsensitivity}) for these detectors. In this context, TREX-DM is a  prototype aimed at testing the feasibility of a Micromegas-based TPC for detecting low-mass WIMPs. The detector can accommodate an active mass of approximately 0.300\,kg of Ar at 10\,bar or approximately 0.160\,kg of Ne at 10\,bar, with an energy threshold below 0.4\,keV, utilizing exclusively radiopure materials. 
\begin{figure}[htbp]
\centering
\includegraphics[width=1\textwidth]{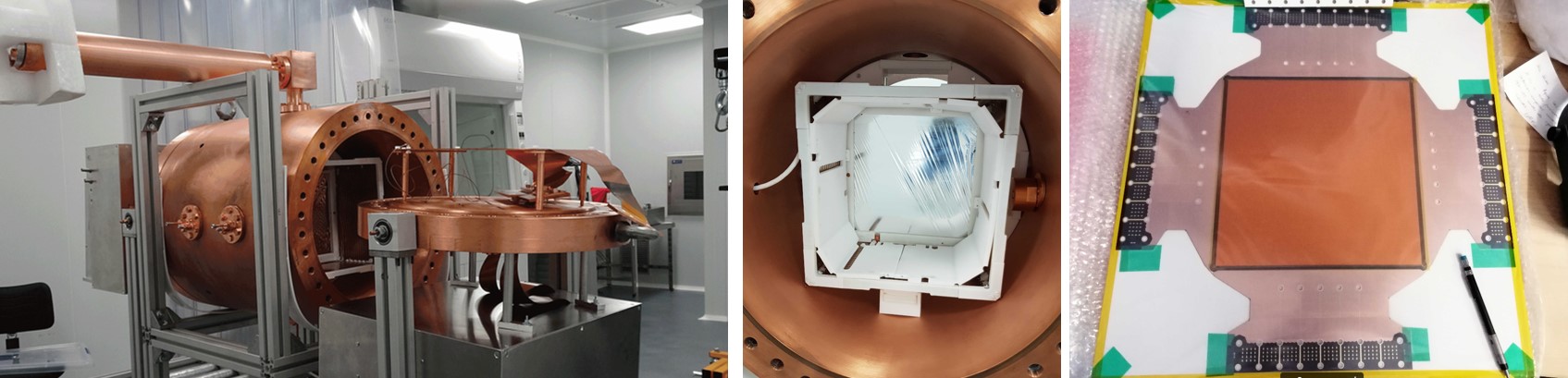}
\caption{
Picture of the TREX-DM detector opened in the LSC clean room, showing: the copper vessel, the field cage inside, and the endcap with the readout plane and the connectors (flat cables) to the electronics  (left); a new PTFE piece inside the  field cage to stop copper florescence (center); and the new microbulk Micromegas readout plane (right).\label{fig:TREXDMMM}}
\end{figure}

TREX-DM is currently installed at the LSC. Its structure includes a copper sleeve, 0.5\,m in diameter and length, which can hold up to 12\,bars, divided into two active volumes by a central cathode. Each volume is surrounded by a field shaper ensuring a uniform drift field, made of copper strips on kapton\textsuperscript{\textregistered} with separating resistors, and the ionization signal is drifted towards the Micromegas readout planes at both endcaps (figure~\ref{fig:TREXDMMM}).

The microbulk Micromegas readouts of TREX-DM have a 25$\times$25\,cm$^2$ active surface, with 256 low capacity strips per axis, linked to four connector prints at the readout sides. It is the largest single piece microbulk Micromegas readout built so far. Efforts to control and further reduce the radioactivity of these planes (especially $^{40}$K contamination) and detaile on a non-commercial Face-to-Face connection concept are described in section \ref{sec:background}.

Additionally, channel tracks and vias are more spaced apart from grounded areas. Regarding pattern and inter-pixel distance, the new design enhances hole quality, thereby improving detector performance in terms of gain and energy resolution.  Increasing the pixel separation from 50\,$\mu$m to 100\,$\mu$m reduces the likelihood of leak currents between pixels in the case of short-circuits with the mesh. 
The acquisition is based on an AGET-based electronic boards allowing an effective energy threshold as low as 400\,eV, although plans to substantially lower it are described in section \ref{sec:lowThreshold}.

A first background model study of TREX-DM was carried out in \cite{Castel2019}, using the REST-for-Physics code \cite{REST2022}. The inputs for simulations were the measured flux of environmental backgrounds at LSC ($\gamma$, n, $\mu$), and the radiopurity data of components taken from measurements from an extensive material screening program to select components (mainly based on Ge $\gamma$  spectrometry at LSC, complemented by GDMS, ICPMS and BiPo-3 measurements -- see \cite{Castel2019} for details). Moreover, the most relevant contributions were identified:  the copper vessel, activated cosmogenically after spending a few years at sea level, which could be mitigated by constructing a new vessel, and the $^{40}$K activity in the Micromegas readout. However, the effects of radon and radon-induced activity were underestimated (see subsection \ref{sec:RadonMitigation}). Contributions from muons and environmental neutrons were estimated under control in simulated conditions, thanks to background rejection capabilities and shielding. Results of the background model for Ar+1\%\,isobutane and Ne+2\%\,isobutane mixtures at 10\,bar
suggested very competitive values of a few counts/keV/kg/day.

\section{Progress in lowering backgrounds} \label{sec:background}

Background events in Micromegas detectors, as in all rare-event searches, come both from intrinsic radioactive contaminants present in the detector components, and from external radiation. General strategies to monitor and control these background sources, typically scrupulous screening of detector components, or shielding techniques, will not be reviewed in this paper. We will focus on a few particular topics that are of specific concern for Micromegas systems like the ones here studied. 

First of all, we will review the efforts to quantify and improve the intrinsic radiopurity of the microbulk Micromegas themselves (subsection~\ref{sec:intrinsic}), a question that is central in the motivation to use these devices in low-background experiments. As will be seen, at the moment the background induced by its radioactivity is well below the experimental level measured in CAST/IAXO and TREX-DM. The second topic regards the DAQ electronics. The complexity of TPCs comes in part from the need to instrument a large number of readout channels. This is typically done using specialized high-channel-density digitizing chips and front-end boards. The Micromegas allow for implementations such that the electronics cards be placed relatively far from the readout, extracting the signal via long flat cables that extend themselves from the very Micromegas plane, thus permitting shielding from typical radioactive electronic components. Nevertheless, placing the electronics closer would bring improvements in threshold and signal quality. We review below (subsection \ref{sec:radiopure_electronics}) the efforts to produce what will be the first TPC high-channel-density DAQ electronics with radiopurity specifications. Our third topic regards background produced by $^{222}$Rn and its progeny (subsection \ref{sec:RadonMitigation}). Despite the fact that this is a well-know problem in many low-background experiments, the way it produces background events in our case is very particular. On one hand, the lightness of the gas target, together with the superb topological information of highly-pixelated readouts, make the typical $\alpha$ or $\beta$ emissions of these isotopes highly suppressible. On the other hand, the absence of self-shielding (due to the lightness of the target mass), as well as of fiducialization in $Z$ direction, allows for some very-low-energy emissions from the $^{222}$Rn progeny at the inner surfaces to directly populate the signal region; a case typically neglected in the literature, because this radiation hardly reaches the sensitive volume in conventional experiments. Finally, we discuss the problem of high-energy cosmic neutrons in surface experiments (subsection \ref{sec:neutrons}), like the case of IAXO. Absent in underground environments (eliminated after a few meters of rock\footnote{Except for a very small amount produced by the few high-energy muons reaching the rock environment of the underground lab.}), they are very penetrating in a surface setup, and require particular shielding designs able to tag the specific topology of the neutron-induced event.

\subsection{Intrinsic radiopurity of Micromegas planes} 
\label{sec:intrinsic}

Traditionally, Micromegas were made by suspending a mesh over anode strips. The mesh was easy to replace for reparations, but it was difficult to maintain a uniform gap over a large area. In the last two decades, new methods have emerged for attaching the mesh to the anode with greater precision. In \textit{bulk} Micromegas \cite{GIOMATARIS2006405}, the mesh is encapsulated within insulating spacers (pillars) attached to the anode, resulting in a unified structure comprising the mesh, pillars, and anode. In \textit{microbulk} detectors~\cite{Microbulk2010}, the mesh, pillars, and readout structure are produced via chemical etching out of a single double-sided copper-clad polyimide (kapton\textsuperscript{\textregistered}) foil, offering not only a low radioactivity budget, but also a more uniform amplification gap compared to other Micromegas types, resulting in better resolution. In parallel with progress with DAQ electronics, these concepts can also lead to a lower energy threshold and a better spatial resolution.

The above-mentioned low radiopurity budget of the microbulks is attributable to the radiopure materials, mainly copper and kapton\textsuperscript{\textregistered}, that they are made of. The radioactivity of Micromegas readout planes was firstly studied in depth and quantified in \cite{Cebrian:2010ta}. Samples representative of the raw materials as well as manufactured readouts were analyzed by germanium spectroscopy at the LSC. On the one hand, two samples were part of fully functional Micromegas detectors: a microbulk readout plane (formerly used in the CAST experiment) and a kapton\textsuperscript{\textregistered} Micromegas anode structure without mesh. On the other hand, two more samples were just raw foils used in the fabrication of microbulk readouts, consisting of kapton\textsuperscript{\textregistered} metallized with copper on one or both sides. The raw materials were confirmed to be very radiopure, bounding their contamination to less than tens of $\mu$Bq/cm$^2$ for the natural $^{238}$U and $^{232}$Th chains and for $^{40}$K. Despite their relevance, these bounds were still relatively modest when expressed in volumetric terms due to the small mass of the samples. Therefore, more sensitive screening techniques and more massive samples were considered for the study of new Micromegas readouts prepared in controlled conditions applying different procedures.

Some of the first samples were prepared for analysis at the BiPo-3 detector \cite{BiPo_detector}. This detector was developed by the SuperNEMO collaboration and operated at the LSC with the aim to measure the extremely low levels of $^{208}$Tl and $^{214}$Bi, produced in the decays of the natural chains of $^{232}$Th and $^{238}$U, in the foils that contain the double-beta decay emitter studied in the experiment. Placing a thin sample foil between two thin layers of scintillators, it is possible to register the BiPo events from the chains by detecting the electron energy deposition in one of the detectors and the delayed alpha signal in the opposite one, reaching sensitivities down to the few $\mu$Bq/kg level. The CAST microbulk Micromegas detector, a Cu-kapton\textsuperscript{\textregistered}-Cu foil and faulty Micromegas circuits produced at CERN were analyzed in the BiPo-3 detector, deriving upper limits at the level of 0.1~$\mu$Bq/cm$^2$, or even less, of $^{214}$Bi and $^{208}$Tl \cite{Castel2019,Irastorza:2015dcb}. Other foils related to the use of Micromegas made of pyralux, kapton\textsuperscript{\textregistered}-epoxy and kapton\textsuperscript{\textregistered}-diamond were also measured in BiPo-3. An updated account of the measurements of these samples with BiPo-3, including additional statistics that is being analysed at present, is under preparation for a forthcoming publication.

\begin{figure}[h]
\centering
\includegraphics[width=0.8\textwidth]{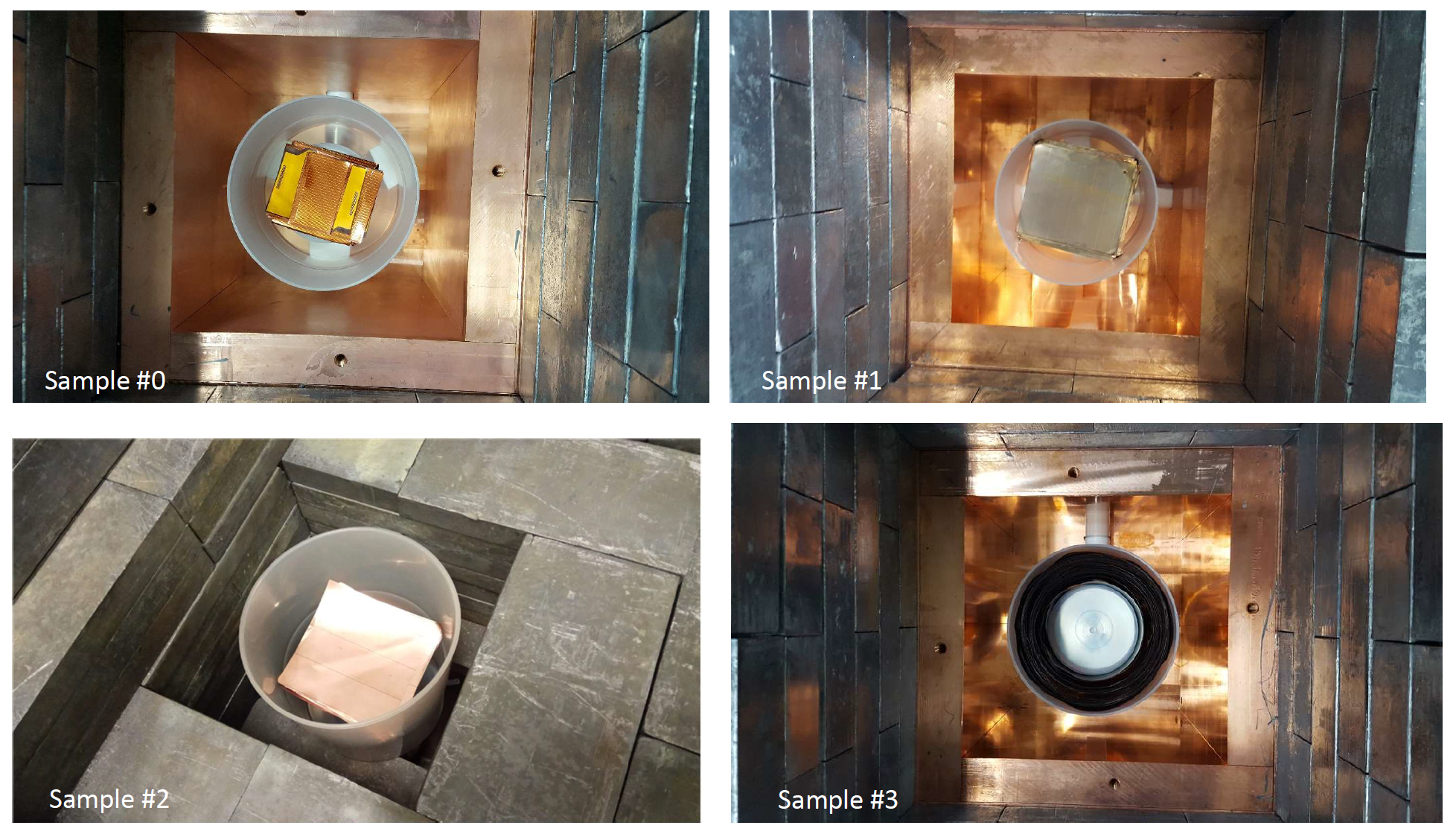}
\caption{Pictures of some of the Micromegas samples prepared at CERN and screened with HPGe detectors at the LSC: the one firstly produced and operated (Sample \#0), after production and different cleaning processes (Samples \#1-2) and a massive sample prepared for sensitive screening (Sample \#3).} \label{FotosGe}
\end{figure}

A dedicated development to quantify, understand and reduce as much as possible the radioactivity of Micromegas readouts, specially the $^{40}$K content, has been undertaken combining controlled production at CERN and a series of radioassays at different steps of production using HPGe detectors at the LSC. A first massive sample (with total surface of 12373\,cm$^2$) was prepared at CERN, consisting of failed GEM glued on kapton\textsuperscript{\textregistered}, as it is done for microbulk Micromegas. The use of tap water during the cleaning process was shown to introduce a relevant uranium contamination, quantifying activities of the upper parts of the $^{238}$U and $^{235}$U chains. Then, new samples (with total surfaces of 7000\,cm$^2$ and 51734.5\,cm$^2$) of microbulk-equivalent Micromegas were prepared at CERN. As baths containing potassium are typically used during Micromegas production (in the kapton\textsuperscript{\textregistered} etching step and in the cleaning step) a new bath was applied trying to neutralise the potassium permanganate. Tap water was avoided and only deionized water was used. Several long measurements (around two months each) using different ultra-low background HPGe detectors at the LSC were carried out during the whole process applied on these samples to quantify activity of the natural chains of $^{238}$U, $^{235}$U and $^{232}$Th as well as of primordial $^{40}$K by gamma spectroscopy, and the first results obtained can be found at \cite{Castel2019}. The cleaning procedure has been shown to be effective, with the lowest $^{40}$K activity quantified being (0.102$\pm$0.030)\,$\mu$Bq/cm$^2$, which means a reduction of a factor $\sim$34 with respect to the value quantified in the first analyzed sample. More recently, a study of the radiopurity of a witness sample (with total surface of 8250\,cm$^2$) of the microbulk Micromegas produced at CERN for TREX-DM has been performed. Figure \ref{FotosGe} shows some the described Micromegas samples. A future publication with all details is in preparation.

The obtained results, combining different techniques to quantify ultra-low radioactivity levels and after specific developments for production at CERN, confirm the suitability of the use of Micromegas as extremely radiopure readouts for rare event searches.

\subsection{Radiopure electronics}
\label{sec:radiopure_electronics}

DAQ electronics for TPCs are relatively complex boards, requiring the handling and digitization of a large number of channels. Due to this, in current detector implementations the front-end boards are placed outside the shielding, relatively far from the detector, thus avoiding radiopurity requirements. The versatility of the microbulk plane design, that can accommodate lateral slaps prolonging the signal strips as extensions of the microbulk foil itself, easily allows for this. However, the possibility of reducing the distance the signal needs to travel until the front-end cards, would help to reach lower electronic noise and lower thresholds, while adding simplicity and robustness in the readout implementation.

\begin{figure}[htbp]
\centering
\includegraphics[width=0.55\textwidth]{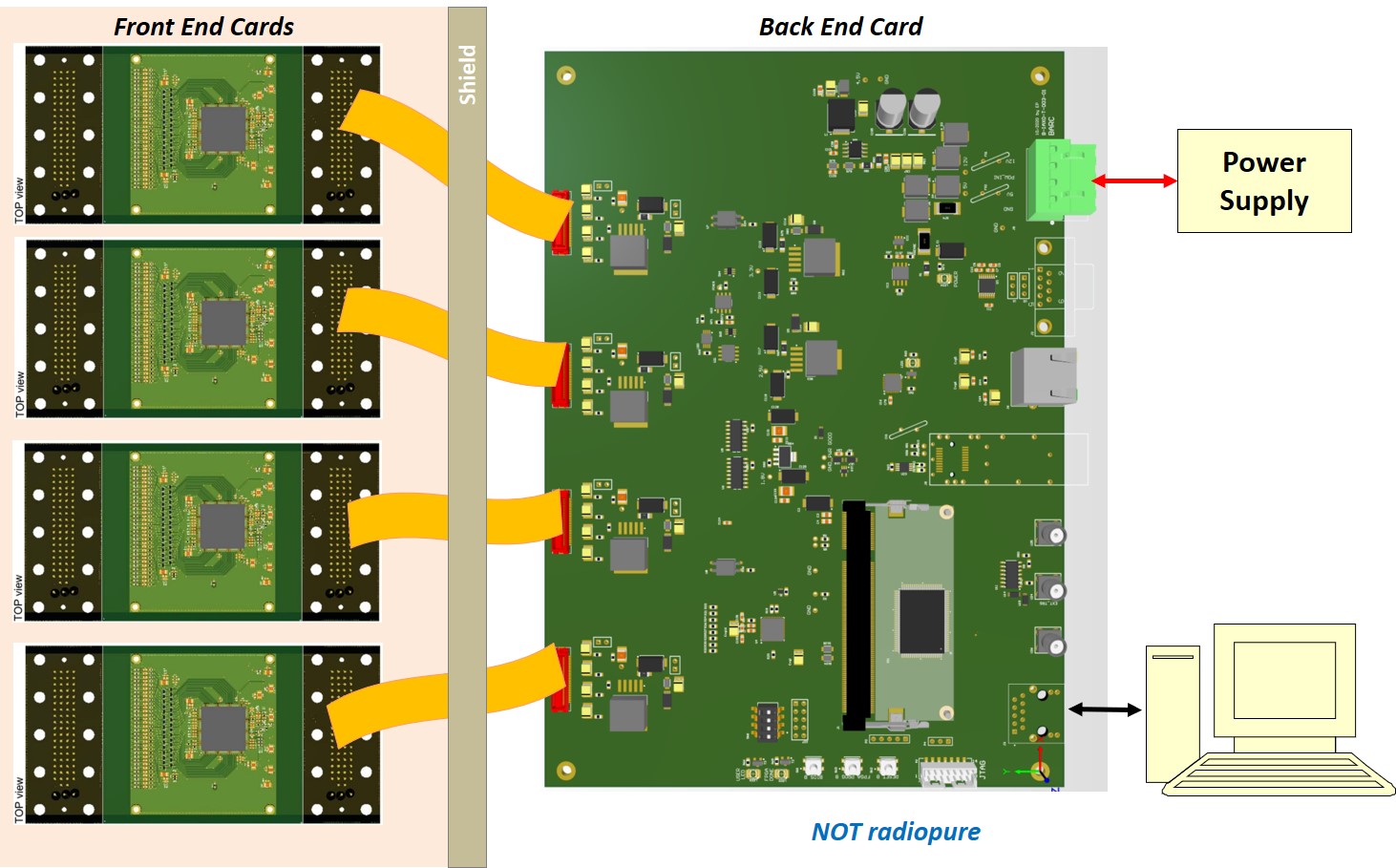}
\includegraphics[width=0.4\textwidth]{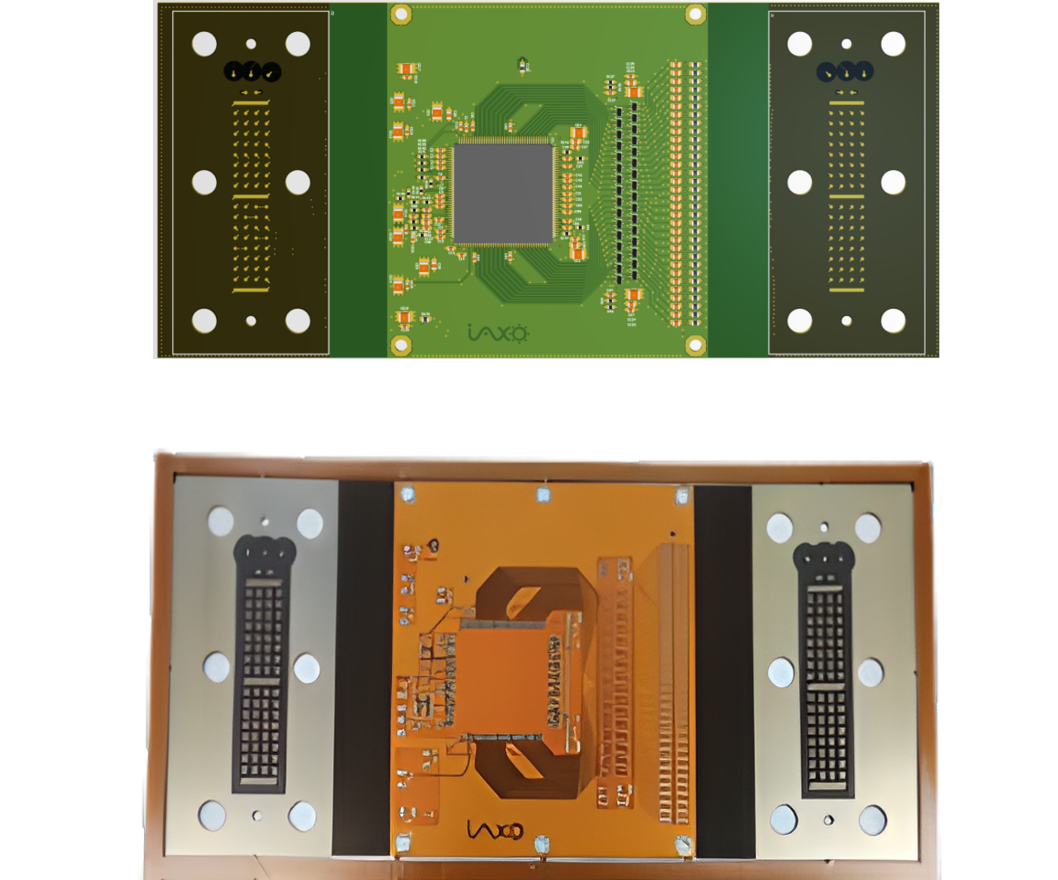}
\caption{Electronics scheme to reduce radioactivity and increase signal-to-noise ratio. The radiopure FECs are close to the detector while the not radiopure BECs are separated by a shield (left). FEC design (right-top) and constructed, ready to be used in IAXO (right-bottom). On the right and left of both images, the pins for FtF connectors, and, in the center, the STAGE chip.}\label{fig:electronics}
\end{figure}

With this goal, the development of a first version of a Micromegas DAQ electronics with radiopure specifications was launched. The previous non-radiopure acquisition board (dubbed ARC) has been divided in two, the new front-end (FEC) and back-end (BEC) cards. The new FEC will be placed very close to the Micromegas and will sustain some radiopurity specifications, while the BEC can be placed relatively far and will contain standard non-radiopure components (see left of figure~\ref{fig:electronics}). Mounted on a clean polyimide substrate, each FEC hosts one state-of-the-art, high-channel-density, auto-triggered STAGE chip from CEA/Saclay, the successor of the AGET chip currently in use in TREX-DM. All components have been screened for radiopurity, and, whenever possible, replaced with a cleaner option. The radioactivity of the chips themselves has also been quantified. A first round of boards have recently been manufactured and are ready for commissioning (see right of figure~\ref{fig:electronics}). The radioactivity target of this first version of radiopure FEC is below 40\,mBq of $^{232}$Th, 600\,mBq of $^{238}$U and 30\,mBq $^{40}$K, being the ceramic capacitors responsible for most of the contamination.  A posteriori screening of the whole board is still pending. As a reference, the background level expected due to the presence of this FEC close to the detector in IAXO is estimated to be below $10^{-10}$\,counts/keV/cm$^2$/s, well below the target background for BabyIAXO and IAXO. The impact of such electronics for TREX-DM is under study. Preliminary discussions with colleagues from the NEWS-G collaboration are ongoing on the possibility of implementing these boards also in that experiment. To our knowledge, it is the first attempt to produce an electronic board of such complexity with radiopure specifications. More details of this effort will be given in a future publication.

Another related problem is the need of connecting the large number of signals typical of TPCs, in a radiopure way. Although microbulks can be designed with extensions bringing the signals relatively far from the detector, thus reducing the need of radiopure connectors, the aforementioned effort of placing the FEC closer would also require new connection solutions. A variety of high-density commercial connectors exist, but, despite substantial search~\cite{Aznar_2013} none of them shows satisfactory radiopure specifications. To solve this issue, the concept of a Face-to-Face (FtF) connector has been developed and tested at the University of Zaragoza. In a FtF connector the pad array footprint in both flexible (kapton\textsuperscript{\textregistered}) substrates are aligned and press against each other by means of a carefully designed mechanics, so that the connection is done just by contact of the confronted pads, without any soldering or intermediate connecting piece. The pressing mechanics is made of copper and expanded PTFE, both hihgly radiopure. Specifications on pad dimensions and pitch, as well as on the pressing parameters, have been derived after exhaustive prototyping and testing, to assure a high degree of reliability and robustness~\cite{tesis_hector}. 
An example of the FtF footprint ca be seen at the sides of the latest version of TREX-DM microbulk, in figure \ref{fig:TREXDMMM}.

\subsection{Mitigation of the effects of radon and its progeny} \label{sec:RadonMitigation}

$^{222}$Rn and its progeny, along with surface contamination induced by exposure to $^{222}$Rn, can pose challenges in direct DM searches (or rare event experiments in general), where very low background levels are required. $^{222}$Rn is usually one of the leading background sources in most of the experiments (e.g. in XENON1T \cite{xenon1t}), and this has been found to be the case in TREX-DM as well.

$^{222}$Rn is a naturally occurring radioactive isotope present in the primordial $^{238}$U decay chain, and therefore can be found in most materials. Given its gaseous nature and its relatively long half-life (it is the longest lived isotope of Rn, with $t_{1/2}=3.82$\,days), its emanation from inner surfaces of materials can reach the active volume of the detector, where the $\beta$ decays of its progeny (and the subsequent low-energy electrons and X-rays) can induce signal-like events, making it hard to eliminate them by means of fiducialisation or offline techniques.
There are essentially two approaches to mitigate the effect of $^{222}$Rn in direct dark matter searches: active removal of $^{222}$Rn from the system, and careful selection of materials with a low $^{238}$U content. While these approaches have been extensively developed in mainstream WIMP experiments, there are two issues that make the $^{222}$Rn problem in gas dark matter detectors like TREX-DM a specific problem.  Firstly, powerful distillation methods to remove $^{222}$Rn~\cite{xenon_rn_distillation} developed by liquid noble detectors cannot be used in experiments like TREX-DM, which runs with gas mixtures (in this case, Ar or Ne with a small percentage of isobutane). This means that we need to rely on \textit{hot} filters or traps to counteract the effect of material emanating $^{222}$Rn, as discussed in the following. Secondly, the high density and full fiducialization capabilities of noble liquid detectors led to the concept of self-shielding to mitigate the effect of surface contaminations, something that is not possible in the lighter gaseous media.

In order to eliminate $^{222}$Rn from the system, several materials can be used as radon traps, with molecular sieves and activated charcoal being two of the most used ones. $^{222}$Rn, being a chemically inert noble gas, can only be trapped via physical adsorption through van der Waals forces. The strategy is to slow down $^{222}$Rn long enough through adsorption/desorption processes so it decays in the trapping material. Its daughters, being electrically charged, can be indefinitely trapped in the material. In order for the adsorption to work, the trapping material must be highly porous, with a pore size big enough to trap $^{222}$Rn. Also, keeping the $^{222}$Rn trap at low temperatures (typically at dry ice temperature) is preferred, since this prolongs the trapping lifetime \cite{charcoal_modane}. This idea was tested in TREX-DM, without yielding the expected result: likely due to the limitations in lowering the temperature imposed by the quencher in the gas mixture (the boiling point of isobutane is -11.7\,ºC).

Molecular sieves are crystalline materials with regular porous structures of a specific size. These pores are such that molecules or atoms below the diameter of the pore can be adsorbed, while bigger molecules are not captured. This allows for a sharp discrimination in terms of molecular size. On the other hand, activated carbon (or charcoal) has a wider range of pore sizes. It is a form of carbon treated (activated) to have a high degree of porosity, which significantly increases its surface area. Both molecular sieves \cite{Rn_removal_mol_sieves} and activated carbon \cite{charcoal} have been found useful in selectively targeting $^{222}$Rn in gases used in ultra-sensitive rare-event experiments. One of the problems with both is self-emanation of $^{222}$Rn, making it critical to select materials with low $^{238}$U content, and using the minimum quantity needed to achieve the $^{222}$Rn requirements of the experiment. In this respect, low-radioactivity molecular sieves have been developed and tested in the context of rare-event experiments \cite{low_radioact_mol_sieves}.

Apart from actively removing $^{222}$Rn from the system, it is essential to use low-radioactivity materials. One of the main challenges stems from the purifiers used in some gas-based rare-event experiments to keep the humidity and oxygen content in check when operating in recirculation mode (the gas is reused once injected). These oxygen and humidity filters have been demonstrated (both by TREX-DM and other experiments) to emanate $^{222}$Rn in different quantities \cite{filters_ieee}. In general, commercially available filters do not meet the radioactivity requirements, and some experiments such as NEWS-G have developed their own custom-made filters, carefully choosing the filtering materials and the quantities \cite{filters_ieee}. One solution to avoid this problem altogether is to continuously flow and dispose of the target gas. A slight variation of this alternative is the semi-sealed open-loop approach, and is especially suited in those cases where the target gas is expensive or difficult to acquire. It involves selecting the minimum flow rate needed to compensate for the outgassing and leaks of the detector vessel, thus keeping the gas clean. If the leak-tightness of the experiment is good, as is the case of TREX-DM, very low flow rates ($<$ 1\,l/h) can be employed, resulting in one renovation every few weeks, a situation that resembles a seal-mode scenario. Shifting to this operation mode successfully reduced the background in TREX-DM by an order of magnitude, getting rid of the $^{222}$Rn emanation problem altogether, and effectively achieving a low-energy level of O(80)\,counts/keV/kg/day.

In addition to volumetric $^{222}$Rn, surface contamination from long exposure to atmospheric $^{222}$Rn can be a potential background source. In particular, deposition of long-lived $^{210}$Pb ($t_{1/2}$=$22.3$\,years), especially on the inner surfaces facing the detector volume, can produce low-energy background due to the emission of X-rays and low-energy electrons. Lighter gas mixtures (Ar- or, especially, Ne-based mixtures), such as the ones used in TREX-DM, are more affected by these low-energy emissions of the decay chain, whereas these emissions are often neglected in experiments using heavier gas mixtures (such as Xe, that provides better self-shielding) and/or dual-phase TPCs, which allow for 3D fiducialisation.

Again, there are two ways of tackling the problem: active removal of the contamination, and screening and protection of the materials used. As for the first approach, removing a thin layer ($\sim$ few mm) of the material, either mechanically or by chemical means (e.g. using nitric acid) has been proved to be successful in getting rid of surface $^{210}$Pb \cite{surface_cont_ptfe}. However, when the contamination directly affects elements where the first approach is not feasible (e.g. the surface of the detectors themselves, or a very thin ($\sim \mu$m) surface such as the mylar\textsuperscript{\textregistered} cathode used in TREX-DM), one has to resort to the use of materials with low intrinsic $^{210}$Pb content, as well as to isolating all the materials as carefully as possible. At the moment, this background is considered the limiting component in TREX-DM, and efforts are ongoing to reduce it. The success of these efforts will determine the sensitivity of future data taking campaigns, something that is discussed in section~\ref{sec:TREXDMsensitivity}.

Given that both $^{222}$Rn and $^{210}$Pb are $\alpha$-emitters, mapping the rate of these particles in special low-gain runs is a good way to monitor the presence of these contaminations in the detector. We have experimentally established an approximately one-to-one correspondence between reduction of high-energy alpha rate and background in the WIMP region-of-interest, something qualitatively understood via preliminary simulations. More thorough simulation-based studies are ongoing. In this context, a high-sensitivity detector that exploits the capabilities of the Micromegas detectors for the reconstruction of alpha particle tracks, dubbed AlphaCAMM (Alpha CAMera Micromegas)~\cite{alphacamm}, has been recently developed and is currently under commissioning. AlphaCAMM will be able to screen flat $\alpha$-emitting samples with high sensitivity, and is expected to become a crucial tool in the quest to mitigate the $^{210}$Pb background in TREX-DM and other experiments, as further discussed later on (subsection~\ref{sec:further}).

\subsection{Cosmic neutrons for surface operation in IAXO}  \label{sec:neutrons}

Cosmic-rays play a significant role in generating background events in Micromegas setups situated at sea level, as it is the case of axion helioscopes. They can penetrate the detector shielding, ionizing gas directly, and may also interact within the shielding, producing secondary fluorescence in the detector's innermost regions, affecting, thus, the lower energy spectrum.

In the framework of the CAST experiment, measurements were carried out at both the University of Zaragoza and the LSC to assess the impact of cosmic muons on the background levels of Micromegas (see subsection \ref{sec:further}) . Studies at surface level with the CAST detector, partially vetoed, showed background levels \cite{nature-2017} an order of magnitude higher compared to measurements carried out at the LSC\cite{Tomas_2012}. Only the penetrating muons reach this depth, although their flux is reduced by 5 orders of magnitude compared to that on the surface.  
Even if the muon-tagging efficiency of vetoes is close to 99\%, there is still an irreducible contribution at surface, attributed to cosmic neutrons, a hypothesis supported by simulations.

A first simulation-based background model for IAXO-D0 was presented in 2019 in \cite{ERuiz}. The background contribution due to intrinsic contaminations of the detector materials and components was bounded below $<10^{-7}$\,counts/keV/cm$^2$/s~\cite{CMargalejo}.  As anticipated, cosmic rays make the most substantial contribution to the background. Among these, cosmic muons generate the highest number of raw background events, approximately 7.2$\times 10^{-4}$\,counts/keV/cm$^2$/s , followed by neutron-induced events, which are an order of magnitude lower at around 1.2$\times 10^{-5}$\,counts/keV/cm$^2$/s. Cosmic gamma rays contribute minimally to the raw background, below 6$\times 10^{-7}$\,counts/keV/cm$^2$/s.
Yet, following the application of discrimination criteria based on the ionization topology in the Micromegas volume, the contributions from all three sources are reduced to a comparable order of magnitude, each approximately a few times $10^{-7}$\,counts/keV/cm$^2$/s, and the combined contribution from cosmic rays (muons, gamma rays, and neutrons) after discrimination is less than 1.6$\times 10^{-6}$\,counts/keV/cm$^2$/s. This result is compatible with previous background assessments focused only on cosmic muons in the CAST-Micromegas setups, without considering the presence of cosmic neutrons. However, after veto tagging, the simulated muon contribution drops to values below $10^{-8}$\,counts/keV/cm$^2$/s, lacking relevance with respect to the neutron contribution, which still remains around 5$\times 10^{-7}$\,counts/keV/cm$^2$/s. According to this simulation-based evidence, cosmic neutrons-induced events could be the dominant source of background of current type of IAXO-like Micromegas detectors, if not properly vetoed with specific neutron-sensitive active shielding.

\begin{figure}[h]
		\centering
     \includegraphics[width=0.355\textwidth]{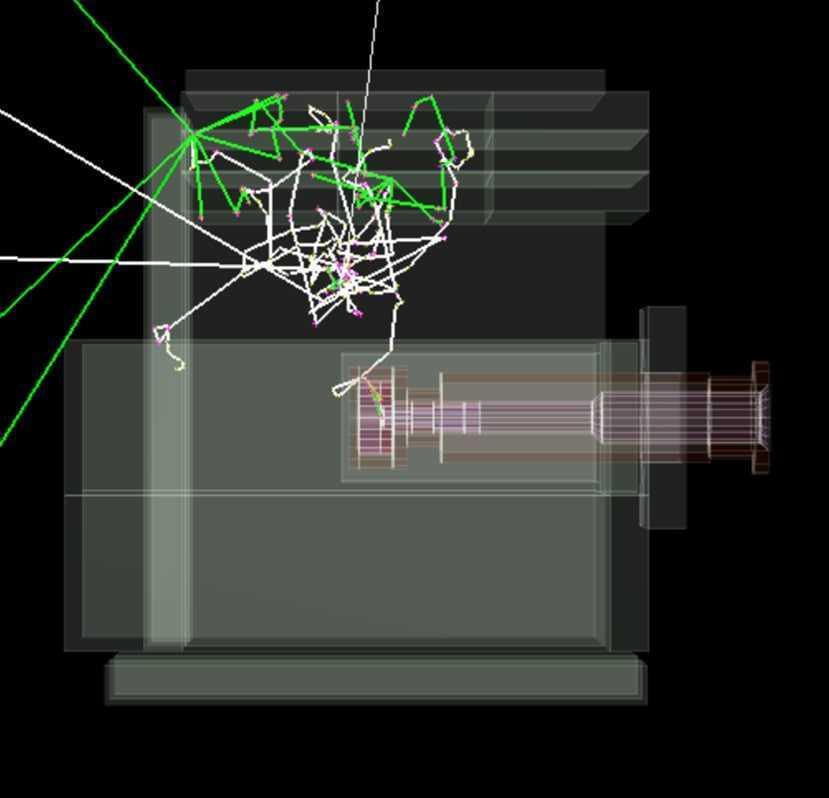}
     \includegraphics[width=0.45\textwidth]{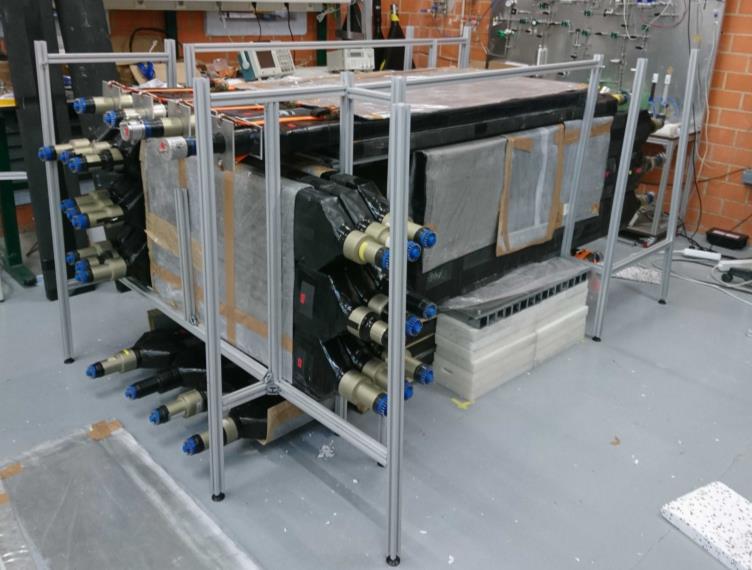}
		\caption{Simulation of a cosmic neutron interaction in the IAXO-D0 setup (left). Picture of the IAXO-D0 setup showing the triple-layer veto system (right).}
		\label{fig:veto_sketch}
	\end{figure}

The IAXO-D0 Micromegas prototype detector has been installed in the IAXOlab of the University of Zaragoza, with the objectives of taking background data, proving simulations, and optimizing its performance under various conditions.
To minimize the effect of natural radiation, the detector has been shielded with 20\,cm of lead and surrounded by a 4$\pi$ muon veto. The analysis shows that the veto system  results in a decrease of the background level 
in the central region of the detector and in the energy region of interest (RoI) for solar axions ([2-7]\,keV) down to 9.8$\times 10^{-7}$\,counts/keV/cm$^2$/s. This number is still higher than what was achieved underground, supporting the cosmic neutron contribution hypothesis. Further efforts have been undertaken to differentiate neutron-induced background, implementing a veto system with 3 layers of  plastic scintillator panels (figure \ref{fig:veto_sketch}) and cadmium sheets surrounding each panel.

Dedicated simulations for this setup indicate that primary neutrons interacting within the setup generate secondary neutrons, which, when captured in cadmium sheets, produce gamma rays detectable in the scintillators. Consequently, neutron events typically result in multiple signals in the veto channels, occurring over several tens of microseconds \cite{Altenmueller-2024}. These simulations have helped to understand the neutron background of our detector; dedicated publications are being prepared with all the details. 

The results of the analysis of the IAXO-D0 data show the lowest background level ever achieved at surface level with this type of detector: $8.6 \times 10^{-7}$\,counts/keV/cm$^2$/s with an efficiency of 79\% for calibration events. The identification of cosmic-neutron-induced signatures in the veto system, allowed to reduce the background level by a 13\% in the energy RoI, with respect to the background level obtained without this handle (but with all other conventional background discrimination criteria). A paper on new background studies, including intrinsic and environmental contamination, as well as veto effects on cosmics, is in preparation.

\subsection{Roadmap towards lower levels} 
\label{sec:further}

\begin{figure}[h]
		\centering\includegraphics[width=0.6\textwidth]{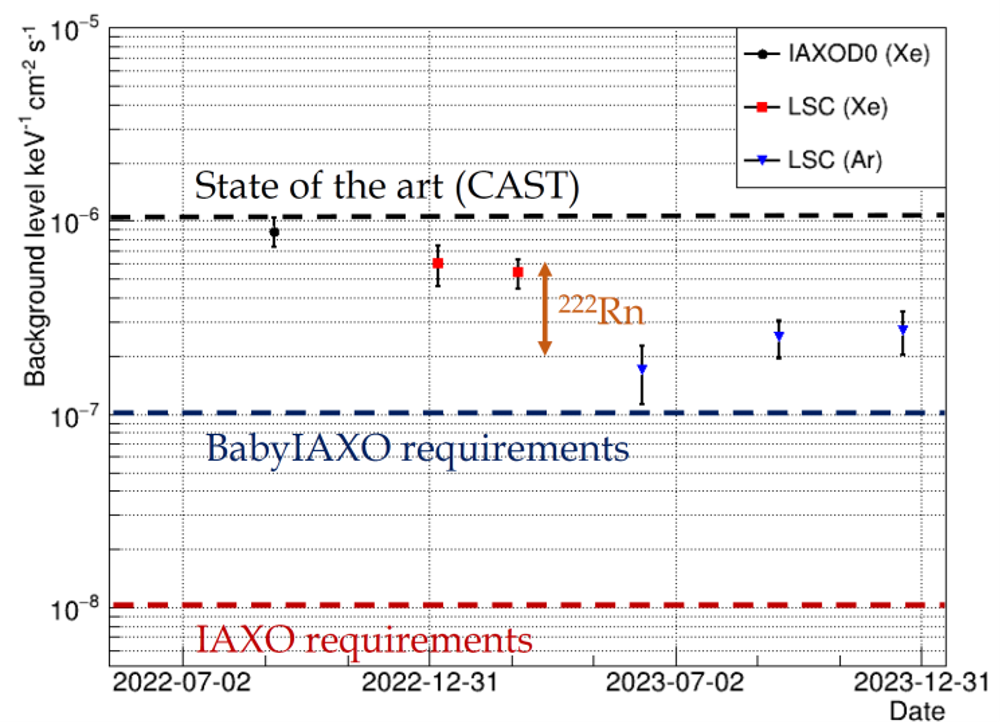}
		\caption{Background level evolution of IAXO-D0 and IAXO-D1. The black dot shows background level reached by IAXO-D1 at surface level. The red squares represent IAXO-D1 background levels at the LSC with Xe-Ne, while the blue triangles show the evolution of IAXO-D1 background levels with Ar.}
		\label{fig:LowBckLevels}
	\end{figure}

Even though the measurement just described represents the lowest background event rate ever achieved at above-ground level with any detector in the IAXO RoI, it still falls short of the levels measured with similar detectors underground. Another prototype, known as IAXO-D1, has been installed underground at the LSC. IAXO-D1 serves two primary objectives: firstly, to compare its data with IAXO-D0, installed on the surface, in order to study the impact of cosmic rays; and secondly, to replicate the background level, which was previously measured, below $2\times10^{-7}$\,counts/keV/cm$^2$/s in the [2-7]\,keV range \cite{ATomas}, in a previous underground setup. This setup was a replica of a CAST microbulk detector, shielded with  10\,cm of external lead and  2.5\,cm of an inner copper layer, and fluxed with N$_2$ to prevent radon intrusion. The IAXO-D1 set-up introduces a new design where new Micromegas detectors have been built to higher standards of radiopurity, with a passive shielding consisting of 20 cm of lead shielding and 1-2.5\,cm of internal copper. IAXO-D1 was installed and commissioned at the LSC during 2022 and has been continuously taking data in different configurations during 2023. Preliminary results (see figure \ref{fig:LowBckLevels}) show background levels of $(5.5\pm1.0)\times 10^{-7}$\,counts/keV/cm$^2$/s with a Xe-Ne mixture, while an even lower background level of about $(1.7\pm0.5)\times 10^{-7}$\,counts/keV/cm$^2$/s  is obtained with Ar. This unexpected result (Xe was preferred a priori to avoid the $^{37}$Ar contamination present in natural Ar gas), may indicate the presence of radon in the Xe-Ne gas mixture.

The most challenging issue in reducing the Micromegas background in axion searches is the tagging of cosmic ray induced neutrons. As explained in \ref{sec:neutrons}, the background at surface level may be dominated by cosmic ray induced neutrons, resulting in nuclear recoils with a similar topology to X-ray events, and therefore cannot be distinguished by event topology. Therefore, the implementation of a highly efficient active veto with neutron tagging capabilities could lead to a background reduction of the Micromegas detectors to their intrinsic level. A new line of research has recently been started to develop a passive and active veto based on the novel LiquidO technique \cite{LiquidO:2019mxd}. It consists of a heavily-doped opaque liquid scintillator with highly granular readout and 3D topological capabilities. On the other hand, the intrinsic background level measured at the LSC may still be limited by the $^{39}$Ar isotope present in the Ar mixture used as conversion gas in the Micromegas. A further background reduction is expected by replacing Ar with Xe. However, the use of Xe typically requires gas purification and recirculation, where the removal of the $^{222}$Rn contamination is particularly challenging. As explained in section \ref{sec:RadonMitigation}, different strategies to minimize this contribution are being investigated. The current state-of-the-art of background level using low background techniques in axion searches, together with the measurements carried out at the LSC are shown in figure \ref{fig:LowBckLevels}.

In the case of TREX-DM, the background levels measured in the 2022 campaign were unexpectedly high, of the order of 1000~\,counts/keV/kg/day. The substitution of contaminated components and the $^{222}$Rn mitigation strategy commented in \ref{sec:RadonMitigation} jointly worked to bring it down to the order of 80~\,counts/keV/kg/day. Dedicated studies conclude that the background level is currently limited by $^{222}$Rn contamination and its progeny. 
An exhaustive material radioassay campaign for TREX-DM has been carried out~\cite{trexdm2016,Castel2019}, mainly based on germanium gamma-ray spectrometry but complemented by other techniques like GDMS or ICPMS. However, these techniques are not adequate to measure concentrations of particular isotopes that can produce alpha surface contamination, like $^{210}$Pb or $^{210}$Po. The AlphaCAMM detector will allow to select the proper raw materials to use inside the TREX-DM chamber, to reduce the background from high energy events and therefore the low energy background in the region of interest, and will provide an accurate input for the background model in TREX-DM.


\section{Progress in lowering the energy threshold} \label{sec:lowThreshold}

\begin{figure}[htbp]
\centering
\includegraphics[width=1.00\textwidth]{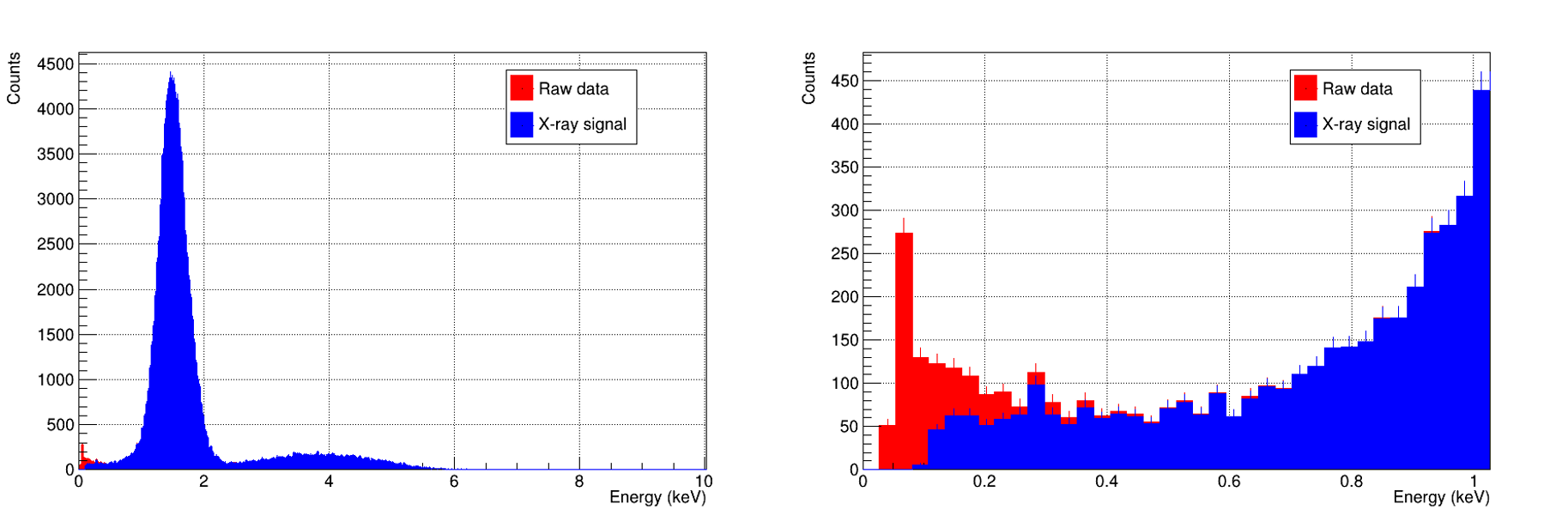}
\caption{The 1.5\,keV X-ray spectrum taken with a microbulk Micromegas  in the X-ray tube of the CAST-lab at CERN  points to an energy threshold of around 0.1\,keV. Full spectrum (left) and a zoomed image for energies below 1\,keV (right). The X-ray signal (in blue) has been  obtained after requiring signals in $X$ and $Y$ directions to raw data (in red).    \label{fig:MM-Threshold}}
\end{figure}

It has already been mentioned that a low energy threshold is an important characteristic for DM experiments and also for helioscopes; in the first case it permits sensitivity to lower WIMP masses, while in the second it opens the possibility to study other physics cases, like the solar axion flux mediated by the axion-electron interaction, a channel that would produce axions of lower energies than the Primakoff solar axions\cite{Barth_2013}. Gas detectors present an intrinsic amplification that could, in principle, allow for very low thresholds, yet in practice, detector features and configurations (readout area, sensor capacitance, electronic noise and general complexity) impose limitations that set a threshold typically limited to a fraction of a keV. A threshold of 450\,eV has been routinely achieved with the last CAST Micromegas detector\cite{SAune_2014} (and even down to 100$\,$eV in particularly optimized calibration runs in the CAST X-ray tube setup at CERN, figure \ref{fig:MM-Threshold} ). Thresholds below 1$\,$keV have been achieved in TREX-DM.

\begin{figure}[htbp]
\centering
\includegraphics[width=0.9\textwidth]{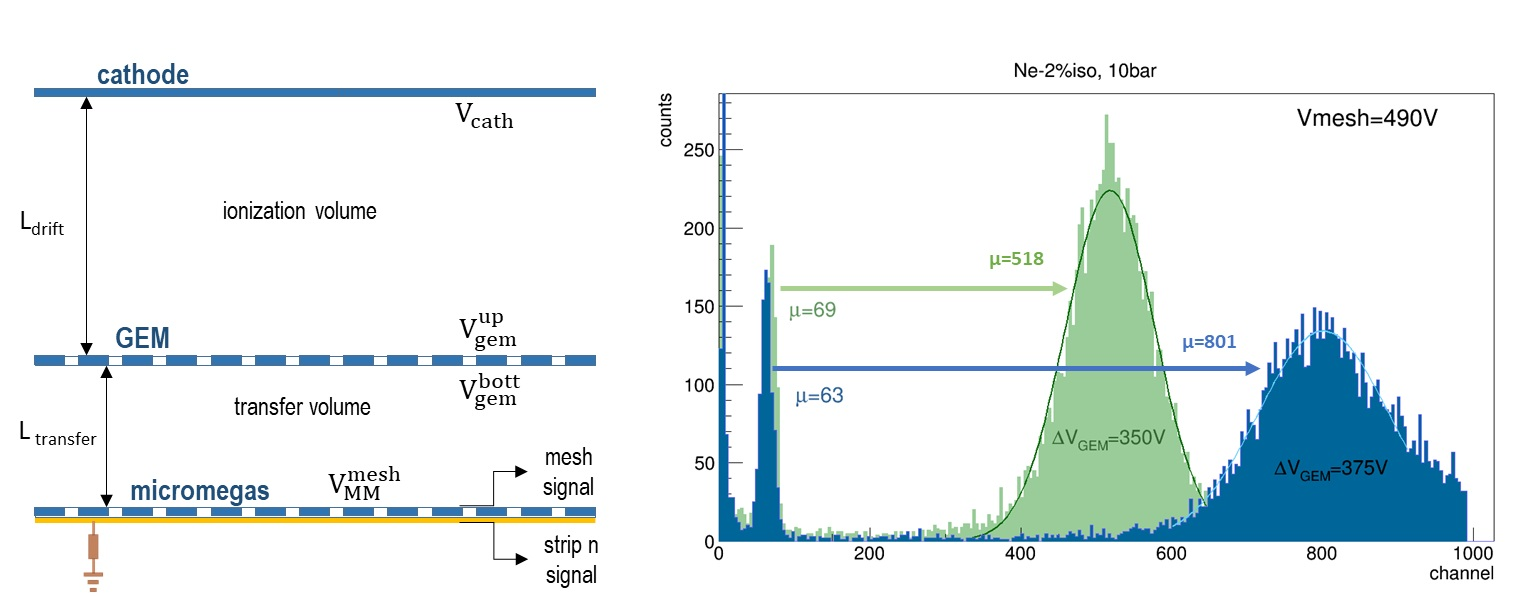}
\caption{A sketch with the idea of a GEM foil suspended above the Micromegas mesh, offering a preamplification stage (left). Energy spectra obtained at two different GEM voltages, where there is a clear separation between the peaks with (centered at $\mu$=518 and $\mu$=801) and without (centered at $\mu$=69 and $\mu$=63) pre-amplification stage. \label{fig:GEM-MM}}
\end{figure}

One promising way to substantially improve the energy threshold is to increase the detector gain by using hybrid readout planes based on Micromegas detectors with a preamplification GEM~(Gas Electron Multiplier) plane (see figure \ref{fig:GEM-MM}) ~\cite{KANE2002}. In the laboratory of the University of Zaragoza (IAXOlab), two setups were prepared: in one, a GEM foil was suspended on top of a small test microbulk Micromegas (2\,cm diameter, non segmented anode) in a chamber where measurements were performed at  pressures of up to 10$\,$bar; in the other, a larger-area GEM was suspended on top of a microbulk Micromegas, spare for the TREX-DM experiment (area of 25\,cm$\times$25\,cm fully segmented anode), in a chamber that cannot withstand more that atmospheric pressure. 
Data were taken in the laboratory (figure \ref{fig:GEM-MM}) using Ar~+~1\%\,isobutane and Ne~+~2\%\,isobutane from 1 to 10\,bar, achieving pre-amplification factors from 12 (Ne~+~2\%\,isobutane at 10\,bar) to 100 (Ar~+~1\%~\,isobutane at 1\,bar). Pre-amplification factors depend on the pressure of the gas, but even in the worst scenario (factor 12 with Ne~+~2\%\,isobutane at 10\,bar) the GEM-Micromegas system could potentially reach energy thresholds down to single-ionization electron. 
More systematic measurements of this preamplification stage are ongoing for different gas mixtures and different detector parameters, and will be published in a dedicated paper. In any case, the current results are already very promising and a program to implement this preamplification stage in TREX-DM is being carried out.

\section{Other technical progress: robustness, ease of construction, scalability} \label{sec:others}

One of the big challenges that many of the rare-event searches face, is the possibility to scale up to bigger volumes. This holds mainly for the WIMP searches, in contrast to the solar axion searches, where the dimension of the detectors can be kept small (thanks to the focusing  of the signal by the X-ray optics). When looking for WIMPs, one needs to maximize the amount of target material offered for interaction, which leads to an increase to much higher volumes and/or pressures of operation, particularly in the case of gas detectors, such is TREX-DM. The increase in volume brings an increase in the active surface of the readout planes, which can be complicated, as the construction of big surfaces entails technical challenges. 

Micromegas detectors of the first generation as big as 40$\times$40\,cm$^2$ were built in the early 2000s \cite{THERS2001133, BOUCHEZ2007425}. On the other hand, several experiments with needs of big surfaces, like the T2K experiment, the MAMMA project of ATLAS or CLAS-12, employ big surfaces of bulk Micromegas of up to 3\,m$^2$ in the case of the modules of the Atlas Muon Spectrometer \cite{Atlas2022}. However, as mentioned above, it is the microbulk Micromegas that is most interesting for the rare-event searches, because of the low radioactivity levels it presents.

\begin{figure}[htbp]
\centering
\includegraphics[width=0.9\textwidth]{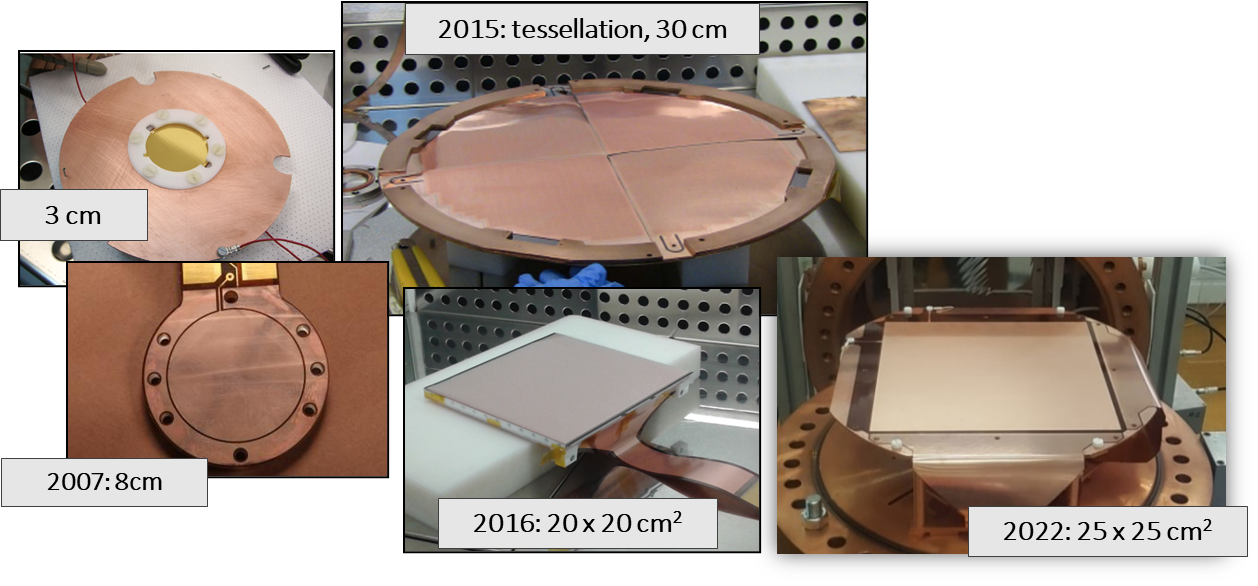}
\caption{Evolution of Micromegas plane sizes over the years: from a radius of 3\,cm to the TREX-DM readout size of 25$\times$25\,cm$^2$ \label{fig:MMscaling}}
\end{figure}

\begin{figure}[h]
		\centering
\includegraphics[width=0.8\textwidth]{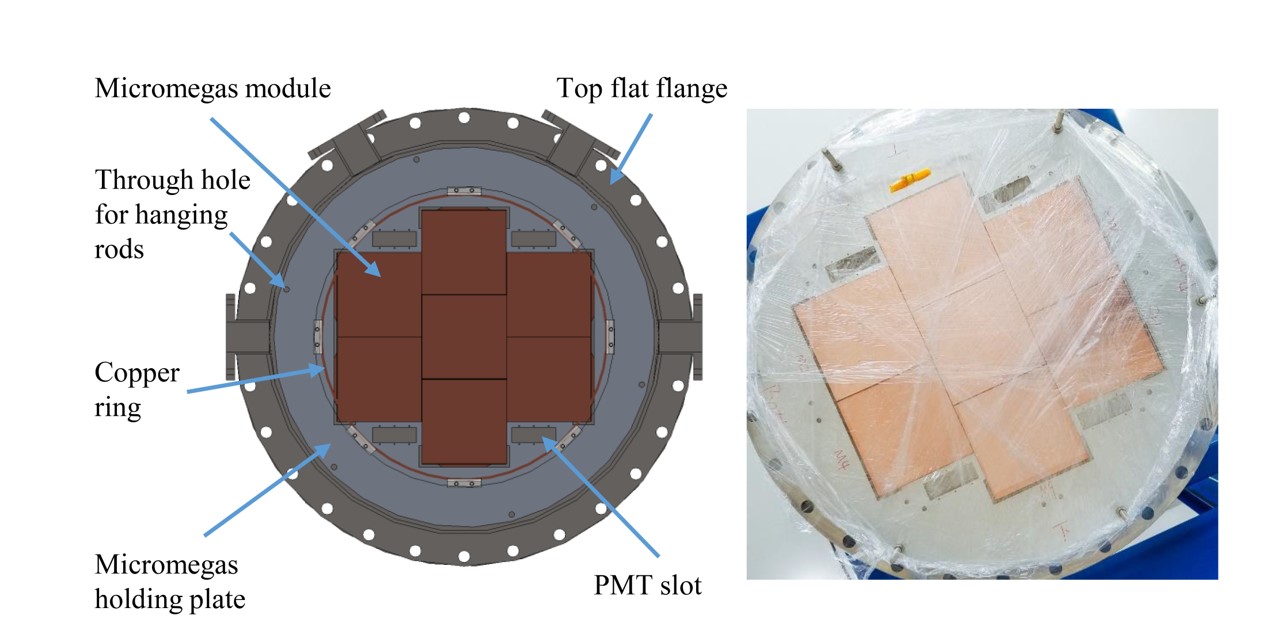}
		\caption{Design and charge readout plane with microbulk Micromegas modules in a  2-3-2 configuration. Adapted from  \cite{Lin_2018}}
		\label{fig:MM_PANDAX}
	\end{figure}

With the help of the CERN Micro Pattern Technology 
(MPT) workshop, and since the invention of the microbulk Micromegas, we have been involved in the production and testing of increasingly bigger surfaces of this particular type of the Micromegas family. Figure \ref{fig:MMscaling} shows a series of pictures with the evolution in active area of the microbulk planes. The infrastructure employed for the fabrication of the microbulks, along with the pressings to accommodate the two-layer anode readout, restrict the safe construction to planes with radii of the order of 20$\,$cm to 25$\,$cm. The largest area microbulk Micromegas so far are the ones built for the TREX-DM experiment (see figure \ref{fig:TREXDMMM}).  In order to cover larger areas, a dead-zone-less tesselation concept was developed in the context of the PandaX-III experiment, including an electrode around each module as electron reintegration system (or \textit{rim} electrode), as well as effective radiopure extraction of signal channels of each module, via extension of the readout planes similar to the ones already shown above for the TREX-DM module. A fist prototype~\cite{Lin_2018}, including a set of 7 microbulk Micromegas modules, each with an active area of 20$\times$20\,cm$^2$,  arranged in a 2-3-2 configuration to cover a total area of 0.3$\,$m$^2$, successfully tested the concept, and is shown in figure \ref{fig:MM_PANDAX}. Based on the tessellation idea, a TPC that will house 140\,kg of enriched Xenon with a reading plane formed by 52 Micromegas modules is currently being commissioned in the PandaX-III experiment~\cite{Zhang2023}.

Another ongoing project is the extension of segmented-mesh microbulk technology to large area detectors for rare event searches. In a segmented-mesh micromegas, in addition to the anode being segmented into strips, the mesh is also divided in several sections, each electrode providing thus one dimension, $X$ or $Y$. This technology \cite{GERALIS:2015bO}, successfully applied to the detection of neutrons for beam profile monitoring \cite{DIAKAKI201846}, can hold certain advantages over standard microbulk designs. One important advantage would be the mitigation of the risk of short circuits between the mesh and strips, thereby eliminating dead channels; another, resolving issues related to low-energy events through a full $XY$ identification. This latter aspect is important to improve the quality of event information for events close to the threshold, and thus of interest for experiments like TREX-DM. The project aims not only to demonstrate the stable operation of large area detectors but also to optimise pitch granularity, offering improvements in event reconstruction and topological discrimination for various rare event searches. 

In the near future there are plans to explore the possibility of using a microbulk Micromegas with a resistive coverlay on the anode. The idea of a \emph{resistive} Micromegas originated for applications where the particle flux is high (e.g. \cite{ALEXOPOULOS2011110}), but it has been applied in many other cases. Covering the segmented anode with a resistive material presents several advantages for the operation of the detector, namely  it improves the stability of the detector, it protects it against the damaging effects of sparks and even protects the readout electronics. In the case of rare event searches, one expects that adding a resistive layer would allow a stable operation in high gains for long times.  The purpose of this R\&D is to prove that, and to study if the resistive materials are radiopure enough for  background requirements.

\section{Sensitivity improvement in TREX-DM to low-mass WIMPs}
\label{sec:TREXDMsensitivity}

The absence of a positive signal in mainstream experiments, as introduced in section \ref{sec:DM-searches}, highlights the need for a broader exploration of the parameter space for WIMPs, including lower mass ranges that may have been previously overlooked. 
\mbox{TREX-DM} was built to search for WIMPs with masses below  10\,GeV, using neon as its main gas, or even argon depleted in $^{39}$Ar. The current levels of background~(80~\,counts/keV/kg/day) and energy threshold~(below ~1~keV$_{ee}$), recorded in \mbox{TREX-DM} in 2022, are not sufficiently low in order to reach the sensitivity required to probe unexplored regions of low-mass WIMPs, and must be improved. In this section of the article we comment on the implementation in TREX-DM of some of the techniques discussed in the previous sections and the effect on the sensitivity of the experiment that these are expected to confer. 

\begin{figure}[htbp]
\centering
\includegraphics[width=0.8\textwidth]{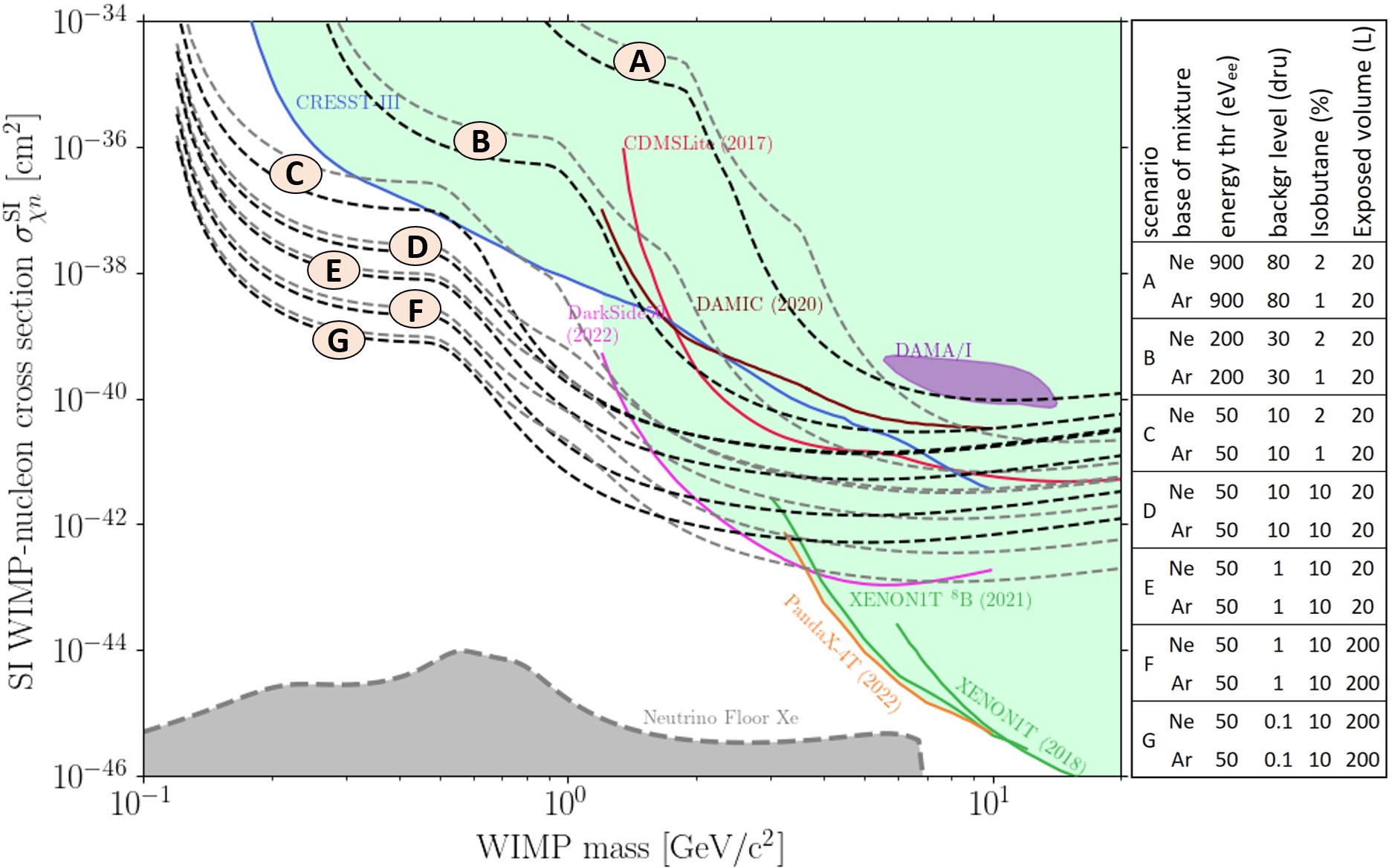}
\caption{WIMP-nucleon cross-section vs. WIMP mass exclusion plot, with current bounds from experiments and \mbox{TREX-DM}, under different conditions for \mbox{TREX-DM} shown in the table at the top-right and 1\,year of exposure time. Each scenario is plotted with \mbox{Ne-based}~(black) and \mbox{Ar-based}~(grey) mixtures. \label{fig:WIMP_ExclusionPlot}}
\end{figure}

Figure~\ref{fig:WIMP_ExclusionPlot} displays multiple sensitivity curves of the experiment for 1 year of exposure time, assuming spin-independent interaction, standard values of the WIMP halo model, astrophysical parameters, and detector parameters discussed below. The depicted curves represent the evolution of the sensitivity  with each projected improvement: scenario $A$ represents the experimental parameters attained during past campaigns, scenarios $B$-$E$ projected for the short to mid-term, and scenarios $F$-$G$ for the long term, with a chamber scaled up by a factor of 10.

The components that enter in direct contact with the active volume of the detector, and therefore would possibly contribute to the measured background levels, are the Micromegas planes, the field cage, the cathode and the gas itself. The microbulk Micromegas installed in \mbox{TREX-DM} in 2018 for physics runs at the LSC~\cite{Aznar2020} were replaced by new microbulk Micromegas in 2022. The new design is more robust from the point of view of the operation stability but also more radiopure, since a novel clean process to reduce the $^{40}$K during the fabrication has been applied, reducing the contamination of $^{40}$K (see subsection \ref{sec:intrinsic}). These measurements estimate the contribution of the Micromegas to the background level of the order of 1\,counts/keV/kg/day.

A comprehensive study of the background origin has led to the conclusion that the principal contribution is due to alpha particles surface contamination, mainly coming from the cathode, likely due to the radon progeny attached to it. Preliminary results point towards an improvement in background level of one order of magnitude, if this component is removed. The cathode consists of a copper frame where a thin foil of aluminized mylar\textsuperscript{\textregistered} \textsuperscript{\textregistered} is tensed. 

In parallel, a fiducial cut of 18$\times$18~cm$^2$ out of the total 25$\times$25~cm$^2$ of the Micromegas has allowed to quantify the high energy events emitted only from the cathode, but there is still an important population of events emitted from the field cage that, as in the case of the cathode, are likely due to the radon progeny. These events can be discriminated by topology studies at the expense of the detector sensitive volume. These two results demand the substitution of the cathode material and of the field cage walls as the first steps to take.

In what regards the energy threshold, the implementation of the preamplification stage discussed in section \ref{sec:lowThreshold} is expected to take place in the short-term: the GEM foils shown in figure \ref{fig:GEM-MM} are ready to be installed.This improvement is expected to bring the energy threshold of TREX-DM, which was 900\,eV during the 2022 data taking campaign at 4$\,$bar, down to 50\,eV, which will have a very important impact on the sensitivity in the low-mass regions below~1\,GeV. Scenarios $B$ and $C$ in figure~\ref{fig:WIMP_ExclusionPlot} represent the expected sensitivity after the successful  reduction of the background level and the energy threshold after the improvements described. Of course, the validity of these scenarios depends on the absence of new, unforeseen background sources at lower energies, something that is not uncommon in the new type of ultralow-threshold WIMP experiments~\footnote{See the EXCESS series of workshops, devoted to low-mass WIMP searches.}.

Moreover, the gas mixture also plays an important role in the sensitivity of TREX-DM. The first sensitivity prospects were estimated for a quantity of isobutane such that the mixture could be classified as non-flammable (1\% in argon and 2\% in neon)~\cite{trexdm2016}. As expected, these prospects were better for Ne- than for Ar-based mixtures (scenarios $A$-$C$ in figure \ref{fig:WIMP_ExclusionPlot}), however the current difficulty in obtaining Ne, favours the use of Ar mixtures. Interestingly, the difference between the two mixtures can be reduced with the increase of isobutane in Ar, as noted when comparing with scenarios $D$-$G$. In addition, the increase of isobutane in the mixtures leads to higher sensitivity to WIMP's masses below~1\,GeV, mainly due to the increase of lower mass nuclei in the target. 

Given this evidence, and taking advantage of the more isolated new site where \mbox{TREX-DM} is installed, the LSC recently accepted the request to operate with flammable mixtures with up to 10\% of isobutane; higher concentrations do not improve significantly the sensitivity and could potentially pose  difficulties to the operation of the detector. 
The use of flammable mixtures requires the application of new safety measurements at the experimental site. These safety measurements are being currently designed and will be implemented in the \mbox{short-mid} term.

\section{Conclusions}\label{sec:conclusions}

We have briefly reported on the application of Micromegas planes in the search for dark matter, particularly the case of low-mass WIMPs with TREX-DM, and solar axions with BabyIAXO. In both cases, Micromegas offer promising prospects in terms of low background and competitive sensitivity to the sought signals. We have stressed various challenges associated with the particular implementation of these readouts, and the ongoing efforts to overcome them. In the case of solar axions, small Micromegas TPCs for the focal point of axion helioscopes are at the moment the most competitive technology. Micromegas prototypes for BabyIAXO have demonstrated background levels below 10$^{-6}$ \,counts/keV/cm$^2$/s in the signal region and, when placed underground, almost down to 10$^{-7}$\,counts/keV/cm$^2$/s, the level targeted for the experiment. Efforts are ongoing to develop active shielding techniques to tag cosmic-neutron-induced events, that are held responsible for preventing reaching this background level also above ground.

In the search for low-mass WIMPs, current TREX-DM sensitivity is hampered by the $^{222}$Rn progeny present on the inner surfaces of the detector, which is believed to dominate the background at low energies. Efforts are ongoing to mitigate this contamination. If successful, the level of radiopurity achieved in the TREX-DM micromegas planes and the rest of the detector parts suggests that very competitive background levels down to 1\,count/keV/kg/day should be achievable. Combining that  with a substantial improvement in energy threshold coming from an additional GEM preamplification stage that is already developed and prepared for commissioning, promising sensitivity close to the single electron, TREX-DM could soon achieve leading sensitivity to WIMPs of few GeV and below. The extent of this assertion will depend very much of the possible presence of unforeseen background sources at such low energies, something not uncommon in this type of experiments.

Future improvements in Micromegas implementations in rare event searches rely on further enhancing radiopurity (both of readout planes or related equipment, like its associated electronics), better robustness from new enhanced layouts (segmented mesh concept, resisitive anode), better energy threshold (via less noisy layouts, e.g. by placing front-end DAQ closer to the readout, and/or via higher gain, with resistive designs or with preamplification stages), optimized gas mixtures (the planned use of Xe in IAXO or the use of higher amount of isobutane in TREX-DM) or, in general, achieving better understanding of limiting background sources beyond the intrinsic ones (the issue of $^{222}$Rn in TREX-DM or the cosmic neutrons in IAXO). In any case, the associated developments continue vigorously and future prospects remain compelling.

\section{Acknowledgements}\label{sec:acknowledgements}

We would like to acknowledge the use of Servicio General de Apoyo a la Investigación-SAI, Universidad de Zaragoza. We acknowledge support from the European Union’s Horizon 2020 research and innovation programme under the European Research Council (ERC) grant agreement ERC-2017-AdG788781 (IAXO+)
and under the Marie Skłodowska-Curie grant agreement No 101026819 (LOBRES). We also acknowledge support from the Agencia Estatal de Investigación (AEI) under grant PID2019-108122GB-C31 funded by MCIN/AEI/10.13039/501100011033, and grant PID2022-137268NB-C51 funded by MCIN/AEI/10.13039/501100011033/FEDER, as well as funds from “European Union NextGenerationEU/PRTR” (Planes complementarios, Programa de Astrof\'isica y F\'isica de Altas Energ\'ias). 
We also acknowledge support from the Agence Nationale de la Recherche (France) ANR-19-CE31-133 0024. 

{\small\bibliographystyle{unsrturl}

\begin{thebibliography}{10}

\bibitem{Blennow_2012}
M.~Blennow, E.~Fernandez Martinez, O.~Mena, J.~Redondo, and P.~Serra.
\newblock Asymmetric dark matter and dark radiation.
\newblock {\em Journal of Cosmology and Astroparticle Physics}, 2012(07):022,
  2012.
\newblock \href {https://doi.org/10.1088/1475-7516/2012/07/022}
  {\path{doi:10.1088/1475-7516/2012/07/022}}.

\bibitem{ZUREK201491}
K.~M. Zurek.
\newblock Asymmetric dark matter: Theories, signatures, and constraints.
\newblock {\em Physics Reports}, 537(3):91--121, 2014.
\newblock \href {https://doi.org/10.1016/j.physrep.2013.12.001}
  {\path{doi:10.1016/j.physrep.2013.12.001}}.

\bibitem{PQ_Axion}
R.~D. Peccei and H.~R. Quinn.
\newblock $\mathrm{CP}$ conservation in the presence of pseudoparticles.
\newblock {\em Phys. Rev. Lett.}, 38:1440--1443, 1977.
\newblock \href {https://doi.org/10.1103/PhysRevLett.38.1440}
  {\path{doi:10.1103/PhysRevLett.38.1440}}.

\bibitem{xenon1t}
XENON Collaboration, E.~Aprile, J.~Aalbers, F.~Agostini, Matteo Alfonsi,
  F.~Amaro, Mouliere Anthony, B.~Antunes, F.~Arneodo, M.~Balata, P.~Barrow,
  L.~Baudis, B.~Bauermeister, M.~L. Benabderrahmane, T.~Berger, Amos Breskin,
  P.~Breur, Andrew Brown, Ethan Brown, and Yi~Zhang.
\newblock {The XENON1T dark matter experiment}.
\newblock {\em The European Physical Journal C}, 77, 2017.
\newblock \href {https://doi.org/10.1140/epjc/s10052-017-5326-3}
  {\path{doi:10.1140/epjc/s10052-017-5326-3}}.

\bibitem{darkside2017}
G.~Zuzel, P.~Agnes, I.~F.~M. Albuquerque, T.~Alexander, A.~K. Alton, D.~M.
  Asner, H.~O. Back, B.~Baldin, K.~Biery, V.~Bocci, G.~Bonfini, W.~Bonivento,
  M.~Bossa, B.~Bottino, A.~Brigatti, J.~Brodsky, F.~Budano, S.~Bussino,
  M.~Cadeddu, L.~Cadonati, M.~Cadoni, F.~Calaprice, N.~Canci, A.~Candela,
  M.~Caravati, M.~Cariello, M.~Carlini, S.~Catalanotti, P.~Cavalcante,
  A.~Chepurnov, C.~Cicalò, A.~G. Cocco, G.~Covone, D.~D'Angelo, M.~D'Incecco,
  S.~Davini, S.~De Cecco, M.~De Deo, M.~De Vincenzi, A.~Derbin, A.~Devoto,
  F.~Di Eusanio, G.~Di Pietro, C.~Dionisi, E.~Edkins, A.~Empl, A.~Fan,
  G.~Fiorillo, K.~Fomenko, G.~Forster, D.~Franco, F.~Gabriele, C.~Galbiati,
  S.~Giagu, C.~Giganti, G.~K. Giovanetti, A.~M. Goretti, F.~Granato, L.~Grandi,
  M.~Gromov, M.~Guan, Y.~Guardincerri, B.~R. Hackett, K.~Herner, D.~Hughes,
  P.~Humble, E.~V. Hungerford, Al. Ianni, An. Ianni, I.~James, T.~N. Johnson,
  C.~Jollet, K.~Keeter, C.~L. Kendziora, G.~Koh, D.~Korablev, G.~Korga,
  A.~Kubankin, X.~Li, M.~Lissia, B.~Loer, P.~Lombardi, G.~Longo, Y.~Ma, I.~N.
  Machulin, A.~Mandarano, S.~M. Mari, J.~Maricic, L.~Marini, C.~J. Martoff,
  A.~Meregaglia, P.~D. Meyers, R.~Milincic, J.~D. Miller, D.~Montanari,
  A.~Monte, B.~J. Mount, V.~N. Muratova, P.~Musico, J.~Napolitano, A.~Navrer
  Agasson, S.~Odrowski, M.~Orsini, F.~Ortica, L.~Pagani, M.~Pallavicini,
  E.~Pantic, S.~Parmeggiano, K.~Pelczar, N.~Pelliccia, A.~Pocar, S.~Pordes,
  D.~A. Pugachev, H.~Qian, K.~Randle, G.~Ranucci, M.~Razeti, A.~Razeto,
  B.~Reinhold, A.~L. Renshaw, M.~Rescigno, Q.~Riffard, A.~Romani, B.~Rossi,
  N.~Rossi, D.~Rountree, D.~Sablone, P.~Saggese, R.~Saldanha, W.~Sands,
  C.~Savarese, B.~Schlitzer, E.~Segreto, D.~A. Semenov, E.~Shields, P.~N.
  Singh, M.~D. Skorokhvatov, O.~Smirnov, A.~Sotnikov, C.~Stanford, Y.~Suvorov,
  R.~Tartaglia, J.~Tatarowicz, G.~Testera, A.~Tonazzo, P.~Trinchese, E.~V.
  Unzhakov, M.~Verducci, A.~Vishneva, B.~Vogelaar, M.~Wada, S.~Walker, H.~Wang,
  Y.~Wang, A.~W. Watson, S.~Westerdale, J.~Wilhelmi, M.~M. Wojcik, Xi. Xiang,
  X.~Xiao, J.~Xu, C.~Yang, A.~Zec, W.~Zhong, and C.~Zhu.
\newblock {The DarkSide Experiment: Present Status and Future}.
\newblock {\em Journal of Physics: Conference Series}, 798(1):012109, 2017.
\newblock \href {https://doi.org/10.1088/1742-6596/798/1/012109}
  {\path{doi:10.1088/1742-6596/798/1/012109}}.

\bibitem{CRESST2019}
A.~H. Abdelhameed, G.~Angloher, P.~Bauer, A.~Bento, E.~Bertoldo, C.~Bucci,
  L.~Canonica, A.~D'Addabbo, X.~Defay, S.~Di~Lorenzo, A.~Erb, F.~v. Feilitzsch,
  S.~Fichtinger, N.~Ferreiro~Iachellini, A.~Fuss, P.~Gorla, D.~Hauff,
  J.~Jochum, A.~Kinast, H.~Kluck, H.~Kraus, A.~Langenk\"amper, M.~Mancuso,
  V.~Mokina, E.~Mondragon, A.~M\"unster, M.~Olmi, T.~Ortmann, C.~Pagliarone,
  L.~Pattavina, F.~Petricca, W.~Potzel, F.~Pr\"obst, F.~Reindl, J.~Rothe,
  K.~Sch\"affner, J.~Schieck, V.~Schipperges, D.~Schmiedmayer, S.~Sch\"onert,
  C.~Schwertner, M.~Stahlberg, L.~Stodolsky, C.~Strandhagen, R.~Strauss,
  C.~T\"urko\ifmmode~\check{g}\else \v{g}\fi{}lu, I.~Usherov, M.~Willers, and
  V.~Zema.
\newblock First results from the cresst-iii low-mass dark matter program.
\newblock {\em Phys. Rev. D}, 100:102002, 2019.
\newblock \href {https://doi.org/10.1103/PhysRevD.100.102002}
  {\path{doi:10.1103/PhysRevD.100.102002}}.

\bibitem{CDMSLite2018}
R.~Agnese, A.~J. Anderson, T.~Aralis, T.~Aramaki, I.~J. Arnquist, W.~Baker,
  D.~Balakishiyeva, D.~Barker, R.~Basu~Thakur, D.~A. Bauer, T.~Binder, M.~A.
  Bowles, P.~L. Brink, R.~Bunker, B.~Cabrera, D.~O. Caldwell, R.~Calkins,
  C.~Cartaro, D.~G. Cerde\~no, Y.~Chang, H.~Chagani, Y.~Chen, J.~Cooley,
  B.~Cornell, P.~Cushman, M.~Daal, P.~C.~F. Di~Stefano, T.~Doughty, L.~Esteban,
  E.~Fascione, E.~Figueroa-Feliciano, M.~Fritts, G.~Gerbier, M.~Ghaith, G.~L.
  Godfrey, S.~R. Golwala, J.~Hall, H.~R. Harris, Z.~Hong, E.~W. Hoppe, L.~Hsu,
  M.~E. Huber, V.~Iyer, D.~Jardin, A.~Jastram, C.~Jena, M.~H. Kelsey,
  A.~Kennedy, A.~Kubik, N.~A. Kurinsky, A.~Leder, B.~Loer, E.~Lopez~Asamar,
  P.~Lukens, D.~MacDonell, R.~Mahapatra, V.~Mandic, N.~Mast, E.~H. Miller,
  N.~Mirabolfathi, R.~A. Moffatt, B.~Mohanty, J.~D. Morales~Mendoza, J.~Nelson,
  J.~L. Orrell, S.~M. Oser, K.~Page, W.~A. Page, R.~Partridge, M.~Pepin,
  M.~Pe\~nalver Martinez, A.~Phipps, S.~Poudel, M.~Pyle, H.~Qiu, W.~Rau,
  P.~Redl, A.~Reisetter, T.~Reynolds, A.~Roberts, A.~E. Robinson, H.~E. Rogers,
  T.~Saab, B.~Sadoulet, J.~Sander, K.~Schneck, R.~W. Schnee, S.~Scorza,
  K.~Senapati, B.~Serfass, D.~Speller, M.~Stein, J.~Street, H.~A. Tanaka,
  D.~Toback, R.~Underwood, A.~N. Villano, B.~von Krosigk, B.~Welliver, J.~S.
  Wilson, M.~J. Wilson, D.~H. Wright, S.~Yellin, J.~J. Yen, B.~A. Young,
  X.~Zhang, and X.~Zhao.
\newblock Low-mass dark matter search with cdmslite.
\newblock {\em Phys. Rev. D}, 97:022002, 2018.
\newblock \href {https://doi.org/10.1103/PhysRevD.97.022002}
  {\path{doi:10.1103/PhysRevD.97.022002}}.

\bibitem{Edelweiss2019}
E.~Armengaud, C.~Augier, A.~Beno\^{\i}t, A.~Benoit, L.~Berg\'e, J.~Billard,
  A.~Broniatowski, P.~Camus, A.~Cazes, M.~Chapellier, F.~Charlieux,
  D.~Ducimeti\`ere, L.~Dumoulin, K.~Eitel, D.~Filosofov, J.~Gascon,
  A.~Giuliani, M.~Gros, M.~De~J\'esus, Y.~Jin, A.~Juillard, M.~Kleifges,
  R.~Maisonobe, S.~Marnieros, D.~Misiak, X.-F. Navick, C.~Nones, E.~Olivieri,
  C.~Oriol, P.~Pari, B.~Paul, D.~Poda, E.~Queguiner, S.~Rozov, V.~Sanglard,
  B.~Siebenborn, L.~Vagneron, M.~Weber, E.~Yakushev, A.~Zolotarova, and B.~J.
  Kavanagh.
\newblock {Searching for low-mass dark matter particles with a massive Ge
  bolometer operated above ground}.
\newblock {\em Phys. Rev. D}, 99:082003, 2019.
\newblock \href {https://doi.org/10.1103/PhysRevD.99.082003}
  {\path{doi:10.1103/PhysRevD.99.082003}}.

\bibitem{CDEX2019}
Z.~Z. Liu, Q.~Yue, L.~T. Yang, K.~J. Kang, Y.~J. Li, H.~T. Wong, M.~Agartioglu,
  H.~P. An, J.~P. Chang, J.~H. Chen, Y.~H. Chen, J.~P. Cheng, Z.~Deng, Q.~Du,
  H.~Gong, X.~Y. Guo, Q.~J. Guo, L.~He, S.~M. He, J.~W. Hu, Q.~D. Hu, H.~X.
  Huang, L.~P. Jia, H.~Jiang, H.~B. Li, H.~Li, J.~M. Li, J.~Li, X.~Li, X.~Q.
  Li, Y.~L. Li, B.~Liao, F.~K. Lin, S.~T. Lin, S.~K. Liu, Y.~D. Liu, Y.~Y. Liu,
  H.~Ma, J.~L. Ma, Y.~C. Mao, J.~H. Ning, H.~Pan, N.~C. Qi, J.~Ren, X.~C. Ruan,
  V.~Sharma, Z.~She, L.~Singh, M.~K. Singh, T.~X. Sun, C.~J. Tang, W.~Y. Tang,
  Y.~Tian, G.~F. Wang, L.~Wang, Q.~Wang, Y.~Wang, Y.~X. Wang, S.~Y. Wu, Y.~C.
  Wu, H.~Y. Xing, Y.~Xu, T.~Xue, N.~Yi, C.~X. Yu, H.~J. Yu, J.~F. Yue, M.~Zeng,
  Z.~Zeng, F.~S. Zhang, M.~G. Zhao, J.~F. Zhou, Z.~Y. Zhou, and J.~J. Zhu.
\newblock {Constraints on Spin-Independent Nucleus Scattering with sub-GeV
  Weakly Interacting Massive Particle Dark Matter from the CDEX-1B Experiment
  at the China Jinping Underground Laboratory}.
\newblock {\em Phys. Rev. Lett.}, 123:161301, Oct 2019.
\newblock \href {https://doi.org/10.1103/PhysRevLett.123.161301}
  {\path{doi:10.1103/PhysRevLett.123.161301}}.

\bibitem{DAMIC2020}
A.~Aguilar-Arevalo, D.~Amidei, D.~Baxter, G.~Cancelo, B.~A.~Cervantes Vergara,
  A.~E. Chavarria, J.~C. D'Olivo, J.~Estrada, F.~Favela-Perez, R.~Ga\"{\i}or,
  Y.~Guardincerri, E.~W. Hoppe, T.~W. Hossbach, B.~Kilminster, I.~Lawson, S.~J.
  Lee, A.~Letessier-Selvon, A.~Matalon, P.~Mitra, C.~T. Overman, A.~Piers,
  P.~Privitera, K.~Ramanathan, J.~Da~Rocha, Y.~Sarkis, M.~Settimo, R.~Smida,
  R.~Thomas, J.~Tiffenberg, M.~Traina, R.~Vilar, and A.~L. Virto.
\newblock Results on low-mass weakly interacting massive particles from an 11
  kg d target exposure of damic at snolab.
\newblock {\em Phys. Rev. Lett.}, 125:241803, 2020.
\newblock \href {https://doi.org/10.1103/PhysRevLett.125.241803}
  {\path{doi:10.1103/PhysRevLett.125.241803}}.

\bibitem{drift}
G.J. Alner, H.~Araujo, A.~Bewick, S.~Burgos, M.J. Carson, J.C. Davies, E.~Daw,
  J.~Dawson, J.~Forbes, T.~Gamble, M.~Garcia, C.~Ghag, M.~Gold, S.~Hollen, R.J.
  Hollingworth, A.S. Howard, J.A. Kirkpatrick, V.A. Kudryavtsev, T.B. Lawson,
  V.~Lebedenko, J.D. Lewin, P.K. Lightfoot, I.~Liubarsky, D.~Loomba,
  R.~Lüscher, J.E. McMillan, B.~Morgan, D.~Muna, A.S. Murphy, G.G. Nicklin,
  S.M. Paling, A.~Petkov, S.J.S. Plank, R.M. Preece, J.J. Quenby, M.~Robinson,
  N.~Sanghi, N.J.T. Smith, P.F. Smith, D.P. Snowden-Ifft, N.J.C. Spooner, T.J.
  Sumner, D.R. Tovey, J.~Turk, E.~Tziaferi, and R.J. Walker.
\newblock {The DRIFT-II dark matter detector: Design and commissioning}.
\newblock {\em Nuclear Instruments and Methods in Physics Research Section A:
  Accelerators, Spectrometers, Detectors and Associated Equipment},
  555(1):173--183, 2005.
\newblock \href {https://doi.org/10.1016/j.nima.2005.09.011}
  {\path{doi:10.1016/j.nima.2005.09.011}}.

\bibitem{mimac}
D.~Santos, J.~Billard, G.~Bosson, J.L. Bouly, O.~Bourrion, Ch. Fourel,
  O.~Guillaudin, F.~Mayet, J.P Richer, A.~Delbart E., Ferrer-Ribas,
  I.~Giomataris, F.J.~Iguaz J.P., Mols, C.~Golabek, and L.~Lebreton.
\newblock {MIMAC : A micro-tpc matrix for directional detection of dark
  matter}.
\newblock {\em EAS Publications Series}, 53:25--31, 2012.
\newblock \href {https://doi.org/10.1051/eas/1253004}
  {\path{doi:10.1051/eas/1253004}}.

\bibitem{cygno}
F.~D. Amaro, E.~Baracchini, L.~Benussi, S.~Bianco, C.~Capoccia, M.~Caponero,
  D.~S. Cardosos, G.~Cavoto, A.~Cortez, I.A. Costa, R.~J. Roque, E.~Dané,
  G.~Dho, F.~Di Giambattista, E.~Di Marco, G.~Grilli di~Cortona, G.~D'Imperio,
  F.~Iacoangel, H.~P.~Lima Júnior, G.~S.~Pinheiro Lopes, A.S.~Lopes Júnior,
  G.~Maccarrone, R.~D.~P. Mano, M.~Marafini, Marcelo R.~R. Gregorio, D.~J.~G.
  Marques, G.~Mazzitelli, A.~G. McLean, A.~Messina, C.M.~Bernardes Monteiro,
  R.~A. Nobrega, I.~F. Pains, E.~Paoletti, L.~Passamonti, S.~Pelosi,
  F.~Petrucci, S.~Piacentini, D.~Piccolo, D.~Pierluigi, D.~Pinci, A.~Prajapati,
  F.~Renga, F.~Rosatelli, A.~Russo, J.~M.~F. dos Santos, G.~Saviano, N.~J.
  Spooner, R.~Tesauro, S.~Tomassini, and Samuele S.~Torelli.
\newblock {The CYGNO Experiment}.
\newblock {\em Instruments}, 6(1), 2022.
\newblock \href {https://doi.org/10.3390/instruments6010006}
  {\path{doi:10.3390/instruments6010006}}.

\bibitem{newsg}
L.~Balogh, C.~Beaufort, A.~Brossard, J.-F. Caron, M.~Chapellier, J.-M.
  Coquillat, E.C. Corcoran, S.~Crawford, A.~Dastgheibi-Fard, Y.~Deng,
  K.~Dering, D.~Durnford, C.~Garrah, G.~Gerbier, I.~Giomataris, G.~Giroux,
  P.~Gorel, M.~Gros, P.~Gros, O.~Guillaudin, E.W. Hoppe, I.~Katsioulas,
  F.~Kelly, P.~Knights, S.~Langrock, P.~Lautridou, I.~Manthos, R.D. Martin,
  J.~Matthews, J.-P. Mols, J.F. Muraz, T.~Neep, K.~Nikolopoulos, P.~O'Brien,
  M.C. Piro, N.~Rowe, D.~Santos, P.~Samuleev, G.~Savvidis, I.~Savvidis,
  F.~Vazquez de~Sola~Fernandez, M.~Vidal, R.~Ward, M.~Zampaolo, and NEWS-G
  collaboration.
\newblock {The NEWS-G detector at SNOLAB}.
\newblock {\em Journal of Instrumentation}, 18(02):T02005, 2023.
\newblock \href {https://doi.org/10.1088/1748-0221/18/02/T02005}
  {\path{doi:10.1088/1748-0221/18/02/T02005}}.

\bibitem{trexdm2016}
F.J. Iguaz, J.G. Garza, F.~Aznar, J.F. Castel, S.~Cebrián, T.~Dafni, J.A.
  García, I.G. Irastorza, A.~Lagraba, G.~Luzón, and A.~Peiró.
\newblock {TREX-DM: a low-background Micromegas-based TPC for low-mass WIMP
  detection}.
\newblock {\em The European Physical Journal C}, 76:529, 2016.
\newblock \href {https://doi.org/10.1140/epjc/s10052-016-4372-6}
  {\path{doi:10.1140/epjc/s10052-016-4372-6}}.

\bibitem{García_2013}
J.~A. García, T.~Dafni, A.~Diago, E~Ferrer-Ribas, J.~Galán, A~Gardikiotis,
  I.~Giomataris, J.~G. Garza, F.~J. Iguaz, I.~G. Irastorza, I.~Ortega,
  T.~Papaevangelou, A.~Rodríguez, A.~Tomás, T.~Vafeiadis, and S.~C. Yildiz.
\newblock {Low-background X-ray detection with Micromegas for axion research}.
\newblock {\em Journal of Physics: Conference Series}, 460(1):012003, 2013.
\newblock \href {https://doi.org/10.1088/1742-6596/460/1/012003}
  {\path{doi:10.1088/1742-6596/460/1/012003}}.

\bibitem{abeln2021conceptual}
A.~Abeln, K.~Altenmüller, S.~Arguedas Cuendis, E.~Armengaud, D.~Attié,
  S.~Aune, S.~Basso, L.~Bergé, B.~Biasuzzi, P.~T. C. Borges~De Sousa, P.~Brun,
  N.~Bykovskiy, D.~Calvet, J.~M. Carmona, J.~F. Castel, S.~Cebrián,
  V.~Chernov, F.~E. Christensen, M.~M. Civitani, C.~Cogollos, T.~Dafní,
  A.~Derbin, K.~Desch, D.~Díez, M.~Dinter, B.~Döbrich, I.~Drachnev,
  A.~Dudarev, L.~Dumoulin, D.~D.~M. Ferreira, E.~Ferrer-Ribas, I.~Fleck,
  J.~Galán, D.~Gascón, L.~Gastaldo, M.~Giannotti, Y.~Giomataris, A.~Giuliani,
  S.~Gninenko, J.~Golm, N.~Golubev, L.~Hagge, J.~Hahn, C.~J. Hailey,
  D.~Hengstler, P.~L. Henriksen, T.~Houdy, R.~Iglesias-Marzoa, F.~J. Iguaz,
  I.~G. Irastorza, C.~Iñiguez, K.~Jakovcic, J.~Kaminski, B.~Kanoute,
  S.~Karstensen, L.~Kravchuk, B.~Lakic, T.~Lasserre, P.~Laurent, O.~Limousin,
  A.~Lindner, M.~Loidl, I.~Lomskaya, G.~López-Alegre, B.~Lubsandorzhiev,
  K.~Ludwig, G.~Luzón, C.~Malbrunot, C.~Margalejo, A.~Marin-Franch,
  S.~Marnieros, F.~Marutzky, J.~Mauricio, Y.~Menesguen, M.~Mentink, S.~Mertens,
  F.~Mescia, J.~Miralda-Escudé, H.~Mirallas, J.~P. Mols, V.~Muratova, X.~F.
  Navick, C.~Nones, A.~Notari, A.~Nozik, L.~Obis, C.~Oriol, F.~Orsini, A.~Ortiz
  de~Solórzano, S.~Oster, H.~P. Pais~Da Silva, V.~Pantuev, T.~Papaevangelou,
  G.~Pareschi, K.~Perez, O.~Pérez, E.~Picatoste, M.~J. Pivovaroff, D.~V. Poda,
  J.~Redondo, A.~Ringwald, M.~Rodrigues, F.~Rueda-Teruel, S.~Rueda-Teruel,
  E.~Ruiz-Choliz, J.~Ruz, E.~O. Saemann, J.~Salvado, T.~Schiffer, S.~Schmidt,
  U.~Schneekloth, M.~Schott, L.~Segui, F.~Tavecchio, H.~H.~J. ten Kate,
  I.~Tkachev, S.~Troitsky, D.~Unger, E.~Unzhakov, N.~Ushakov, J.~K. Vogel,
  D.~Voronin, A.~Weltman, U.~Werthenbach, W.~Wuensch, and A.~Yanes-Díaz.
\newblock {Conceptual Design of BabyIAXO, the intermediate stage towards the
  International Axion Observatory}.
\newblock {\em JHEP05}, 2021.
\newblock \href {https://doi.org/10.1007/JHEP05(2021)137}
  {\path{doi:10.1007/JHEP05(2021)137}}.

\bibitem{MMs}
D.~Attié, S.~Aune, E.~Berthoumieux, F.~Bossù, P.~Colas, A.~Delbart,
  E.~Dupont, E.~Ferrer~Ribas, I.~Giomataris, A.~Glaenzer, H.~Gómez,
  F.~Gunsing, F.~Jambon, F.~Jeanneau, M.~Lehuraux, D.~Neyret, T.~Papaevangelou,
  E.~Pollacco, S.~Procureur, M.~Revolle, P.~Schune, L.~Segui, L.~Sohl,
  M.~Vandenbroucke, and Z.~Wu.
\newblock Current status and future developments of micromegas detectors for
  physics and applications.
\newblock {\em Applied Sciences}, 11(12), 2021.
\newblock \href {https://doi.org/10.3390/app11125362}
  {\path{doi:10.3390/app11125362}}.

\bibitem{Irastorza:2015dcb}
I.G. Irastorza, F.~Aznar, J.~Castel, S.~Cebrián, T.~Dafni, J.~Galán, J.A.
  Garcia, J.G. Garza, H.~Gómez, D.C. Herrera, F.J. Iguaz, G.~Luzon,
  H.~Mirallas, E.~Ruiz, L.~Seguí, and A.~Tomás.
\newblock {{Gaseous time projection chambers for rare event detection: Results
  from the T-REX project. I. Double beta decay}}.
\newblock {\em JCAP}, 1601:033, 2016.
\newblock \href {https://doi.org/10.1088/1475-7516/2016/01/033}
  {\path{doi:10.1088/1475-7516/2016/01/033}}.

\bibitem{Irastorza:2015geo}
I.G. Irastorza, F.~Aznar, J.~Castel, S.~Cebrián, T.~Dafni, J.~Galán, J.A.
  Garcia, J.G. Garza, H.~Gómez, D.C. Herrera, F.J. Iguaz, G.~Luzon,
  H.~Mirallas, E.~Ruiz, L.~Seguí, and A.~Tomás.
\newblock {{Gaseous time projection chambers for rare event detection: Results
  from the T-REX project. II. Dark matter}}.
\newblock {\em JCAP}, 01:034, 2016.
\newblock [Erratum: JCAP 05, E01 (2016)].
\newblock \href {https://doi.org/10.1088/1475-7516/2016/05/E01}
  {\path{doi:10.1088/1475-7516/2016/05/E01}}.

\bibitem{nature-2017}
V.~Anastassopoulos, S.~Aune, K.~Bart, A.~Belov, H.~Bräuninger, G.~Cantatore,
  J.~M. Carmona, J.~F. Castel, F.~Christensen S.~A.~Cetin, M.~Davenport
  J.~I.~Collar, T.~Dafni, T.~A. Decker, A.~Dermenev, K.~Desch,
  C.~Eleftheriadis, G.~Fanourakis, E.~Ferrer-Ribas, H.~Fischer, J.~A. Garcia,
  J.~G.~Garza A.~Gardikiotis~and, E.~N. Gazis, T.~Geralis, I.~Giomataris,
  S.~Gninenko, C.~J. Hailey, M.~D. Hasinoff, D.~H.~H. Hoffmann, F.~J. Iguaz,
  I.~G. Irastorza, A.~Jakobsen, J.~Jacoby, K.~Jakovcic, J.~Kaminski, M.~Karuza
  andN. Kralj, M.~Krcmar, S.~Kostoglou, Ch. Krieger, B.~Lakic, J.~M. Laurent,
  A.~Liolios, A.~Ljubicic, G.~Luzon, M.~Maroudas, L.~Miceli, S.~Neff,
  I.~Ortega, T.~Papaevangelou, K.~Paraschou, M.~J. Pivovaroff, G.~Raffelt,
  M.~Rosu, J.~Ruz, E.~Ruiz Choliz, I.~Savvidis, S.~Schmidt, Y.~K. Semertzidis,
  S.~K. Solanki, L.~Stewart, T.~Vafeiadis, J.~K. Vogel, S.~C. Yildiz, and
  K.~Zioutas.
\newblock {New CAST limit on the axion–photon interaction}.
\newblock {\em Nature Physics}, 13(6):584–590, 2017.
\newblock \href {https://doi.org/10.1038/nphys4109}
  {\path{doi:10.1038/nphys4109}}.

\bibitem{CAST-He3}
M.~Arik, S.~Aune, K.~Barth, A.~Belov, S.~Borghi, H.~Bräuninger, G.~Cantatore,
  J.M. Carmona, S.A. Cetin, J.I. Collar, E.~Da~Riva, T.~Dafni, M.~Davenport,
  C.~Eleftheriadis, N.~Elias, G.~Fanourakis, E.~Ferrer-Ribas, P.~Friedrich,
  J.~Galán, J.A. García, A.~Gardikiotis, J.G. Garza, E.N. Gazis, T.~Geralis,
  E.~Georgiopoulou, I.~Giomataris, S.~Gninenko, H.~Gómez, M.~Gómez~Marzoa,
  E.~Gruber, T.~Guthörl, R.~Hartmann, S.~Hauf, F.~Haug, M.D. Hasinoff, D.H.H.
  Hoffmann, F.J. Iguaz, I.G. Irastorza, J.~Jacoby, K.~Jakovčić, M.~Karuza,
  K.~Königsmann, R.~Kotthaus, M.~Krčmar, M.~Kuster, B.~Lakić, P.M. Lang,
  J.M. Laurent, A.~Liolios, A.~Ljubičić, G.~Luzón, S.~Neff, T.~Niinikoski,
  A.~Nordt, T.~Papaevangelou, M.J. Pivovaroff, G.~Raffelt, H.~Riege,
  A.~Rodríguez, M.~Rosu, J.~Ruz, I.~Savvidis, I.~Shilon, P.S. Silva, S.K.
  Solanki, L.~Stewart, A.~Tomás, M.~Tsagri, K.~van Bibber, T.~Vafeiadis,
  J.~Villar, J.K. Vogel, S.C. Yildiz, and K.~Zioutas.
\newblock {Search for Solar Axions by the CERN Axion Solar Telescope with
  $^3$He Buffer Gas: Closing the Hot Dark Matter Gap}.
\newblock {\em Physical Review Letters}, 112(9), 2014.
\newblock \href {https://doi.org/10.1103/physrevlett.112.091302}
  {\path{doi:10.1103/physrevlett.112.091302}}.

\bibitem{CAST-He4}
M.~Arik, S.~Aune, K.~Barth, A.~Belov, H.~Bräuninger, J.~Bremer, V.~Burwitz,
  G.~Cantatore, J.M. Carmona, S.A. Cetin, J.I. Collar, E.~Da~Riva, T.~Dafni,
  M.~Davenport, A.~Dermenev, C.~Eleftheriadis, N.~Elias, G.~Fanourakis,
  E.~Ferrer-Ribas, J.~Galán, J.A. García, A.~Gardikiotis, J.G. Garza, E.N.
  Gazis, T.~Geralis, E.~Georgiopoulou, I.~Giomataris, S.~Gninenko,
  M.~Gómez~Marzoa, M.D. Hasinoff, D.H.H. Hoffmann, F.J. Iguaz, I.G. Irastorza,
  J.~Jacoby, K.~Jakovčić, M.~Karuza, M.~Kavuk, M.~Krčmar, M.~Kuster,
  B.~Lakić, J.M. Laurent, A.~Liolios, A.~Ljubičić, G.~Luzón, S.~Neff,
  T.~Niinikoski, A.~Nordt, I.~Ortega, T.~Papaevangelou, M.J. Pivovaroff,
  G.~Raffelt, A.~Rodríguez, M.~Rosu, J.~Ruz, I.~Savvidis, I.~Shilon, S.K.
  Solanki, L.~Stewart, A.~Tomás, T.~Vafeiadis, J.~Villar, J.K. Vogel, S.C.
  Yildiz, and K.~Zioutas.
\newblock {New solar axion search using the CERN Axion Solar Telescope with
  $^4$He filling}.
\newblock {\em Physical Review D}, 92(2), 2015.
\newblock \href {https://doi.org/10.1103/physrevd.92.021101}
  {\path{doi:10.1103/physrevd.92.021101}}.

\bibitem{Armengaud_2014}
E.~Armengaud, F.~T. Avignone, M.~Betz, P.~Brax, P.~Brun, G.~Cantatore, J.~M.
  Carmona, G.~P. Carosi, F.~Caspers, S.~Caspi, S.~A. Cetin, D.~Chelouche, F.~E.
  Christensen, A.~Dael, T.~Dafni, M.~Davenport, A.~V. Derbin, K.~Desch,
  A.~Diago, B.~Dölbrich, I.~Dratchnev, A.~Dudarev, C.~Eleftheriadis,
  G.~Fanourakis, E.~Ferrer-Ribas, J.~Galán, J.~A. García, J.~G. Garza,
  T.~Geralis, B.~Gimeno, I.~Giomataris, S.~Gninenko, H.~Gómez,
  D.~González-Díaz, E.~Guendelman, C.~J. Hailey, T.~Hiramatsu, D.~H.~H.
  Hoffmann, D.~Horns, F.~J. Iguaz, I.~G. Irastorza, J.~Isern, K.~Imai, A.~C.
  Jakobsen, J.~Jaeckel, K.~Jakovčić, J.~Kaminski, M.~Kawasaki, M.~Karuza,
  M.~Krčmar, K.~Kousouris, C.~Krieger, B.~Lakić, O.~Limousin, A.~Lindner,
  A.~Liolios, G.~Luzón, S.~Matsuki, V.~N. Muratova, C.~Nones, I.~Ortega,
  T.~Papaevangelou, M.~J. Pivovaroff, G.~Raffelt, J.~Redondo, A.~Ringwald,
  S.~Russenschuck, J.~Ruz, K.~Saikawa, I.~Savvidis, T.~Sekiguchi, Y.~K.
  Semertzidis, I.~Shilon, P.~Sikivie, H.~Silva, H.~ten Kate, A.~Tomas,
  S.~Troitsky, T.~Vafeiadis, K.~van Bibber, P.~Vedrine, J.~A. Villar, J.~K.
  Vogel, L.~Walckiers, A.~Weltman, W.~Wester, S.~C. Yildiz, and K.~Zioutas.
\newblock {Conceptual design of the International Axion Observatory (IAXO)}.
\newblock {\em Journal of Instrumentation}, 9(05):T05002, 2014.
\newblock \href {https://doi.org/10.1088/1748-0221/9/05/T05002}
  {\path{doi:10.1088/1748-0221/9/05/T05002}}.

\bibitem{Armengaud_2019}
E.~Armengaud, D.~Attié, S.~Basso, P.~Brun, N.~Bykovskiy, J.M. Carmona, J.F.
  Castel, S.~Cebrián, M.~Cicoli, M.~Civitani, C.~Cogollos, J.P. Conlon,
  D.~Costa, T.~Dafni, R.~Daido, A.V. Derbin, M.A. Descalle, K.~Desch, I.S.
  Dratchnev, B.~Döbrich, A.~Dudarev, E.~Ferrer-Ribas, I.~Fleck, J.~Galán,
  G.~Galanti, L.~Garrido, D.~Gascon, L.~Gastaldo, C.~Germani, G.~Ghisellini,
  M.~Giannotti, I.~Giomataris, S.~Gninenko, N.~Golubev, R.~Graciani, I.G.
  Irastorza, K.~Jakovčić, J.~Kaminski, M.~Krčmar, C.~Krieger, B.~Lakić,
  T.~Lasserre, P.~Laurent, O.~Limousin, A.~Lindner, I.~Lomskaya,
  B.~Lubsandorzhiev, G.~Luzón, M.~C.~D. Marsh, C.~Margalejo, F.~Mescia,
  M.~Meyer, J.~Miralda-Escudé, H.~Mirallas, V.N. Muratova, X.F. Navick,
  C.~Nones, A.~Notari, A.~Nozik, A.~Ortiz de~Solórzano, V.~Pantuev,
  T.~Papaevangelou, G.~Pareschi, K.~Perez, E.~Picatoste, M.J. Pivovaroff,
  J.~Redondo, A.~Ringwald, M.~Roncadelli, E.~Ruiz-Chóliz, J.~Ruz, K.~Saikawa,
  J.~Salvadó, M.P. Samperiz, T.~Schiffer, S.~Schmidt, U.~Schneekloth,
  M.~Schott, H.~Silva, G.~Tagliaferri, F.~Takahashi, F.~Tavecchio, H.~ten Kate,
  I.~Tkachev, S.~Troitsky, E.~Unzhakov, P.~Vedrine, J.K. Vogel, C.~Weinsheimer,
  A.~Weltman, and W.~Yin.
\newblock {Physics potential of the International Axion Observatory (IAXO)}.
\newblock {\em Journal of Cosmology and Astroparticle Physics}, 2019(06):047,
  2019.
\newblock \href {https://doi.org/10.1088/1475-7516/2019/06/047}
  {\path{doi:10.1088/1475-7516/2019/06/047}}.

\bibitem{NGAH}
I.G. Irastorza, F.T. Avignone, S.~Caspi, J.M. Carmona, T.~Dafni, M.~Davenport,
  A.~Dudarev, G.~Fanourakis, E.~Ferrer-Ribas, J.~Galán, J.A. García,
  T.~Geralis, I.~Giomataris, H.~Gómez, D.H.H. Hoffmann, F.J. Iguaz,
  K.~Jakovčić, M.~Krčmar, B.~Lakić, G.~Luzón, M.~Pivovaroff,
  T.~Papaevangelou, G.~Raffelt, J.~Redondo, A.~Rodríguez, S.~Russenschuck,
  J.~Ruz, I.~Shilon, H.~Ten Kate, A.~Tomás, S.~Troitsky, K.~van Bibber, J.A.
  Villar, J.~Vogel, L.~Walckiers, and K.~Zioutas.
\newblock Towards a new generation axion helioscope.
\newblock {\em Journal of Cosmology and Astroparticle Physics}, 2011(06):013,
  2011.
\newblock \href {https://doi.org/10.1088/1475-7516/2011/06/013}
  {\path{doi:10.1088/1475-7516/2011/06/013}}.

\bibitem{Abbon_2007}
P.~Abbon, S.~Andriamonje, S.~Aune, Th. Dafni, M.~Davenport, E.~Delagnes,
  R.~de~Oliveira, G.~Fanourakis, E.~Ferrer Ribas, J.~Franz, T.~Geralis,
  A.~Giganon, M.~Gros, Y.~Giomataris, I.~G. Irastorza, K.~Kousouris,
  J.~Morales, T.~Papaevangelou, J.~Ruz, K.~Zachariadou, and K.~Zioutas.
\newblock {The Micromegas detector of the CAST experiment}.
\newblock {\em New Journal of Physics}, 9(6):170, 2007.
\newblock \href {https://doi.org/10.1088/1367-2630/9/6/170}
  {\path{doi:10.1088/1367-2630/9/6/170}}.

\bibitem{SAune_2013}
S.~Aune, F.~Aznar, D.~Calvet, T.~Dafni, A.~Diago, F.~Druillole, G.~Fanourakis,
  E.~Ferrer-Ribas, J.~Galán, J.~A. García, A.~Gardikiotis, J.~G. Garza,
  T.~Geralis, I~Giomataris, H.~Gómez, D.~González-Díaz, D.~C. Herrera, F.~J.
  Iguaz, I.~G. Irastorza, D.~Jourde, G.~Luzón, H.~Mirallas, J.~P. Mols,
  T.~Papaevangelou, A.~Rodríguez, L.~Seguí, A.~Tomás, T.~Vafeiadis, and
  S.~C. Yildiz.
\newblock {X-ray detection with Micromegas with background levels below
  $10^{-6}$keV$^{-1}$cm$^{-2}$s$^{-1}$}.
\newblock {\em Journal of Instrumentation}, 8(12):C12042, 2013.
\newblock \href {https://doi.org/10.1088/1748-0221/8/12/C12042}
  {\path{doi:10.1088/1748-0221/8/12/C12042}}.

\bibitem{SAune_2014}
S.~Aune, J.~F. Castel, T.~Dafni, M.~Davenport, G.~Fanourakis, E.~Ferrer-Ribas,
  J.~Galán, J.~A. García, A.~Gardikiotis, T.~Geralis, I.~Giomataris,
  H.~Gómez, J.~G. Garza, D.~C. Herrera, F.~J. Iguaz, I.~G. Irastorza,
  D.~Jourde, G.~Luzón, J.~P. Mols, T.~Papaevangelou, A.~Rodríguez, J.~Ruz,
  L.~Seguí, A.~Tomás, T.~Vafeiadis, and S.~C. Yildiz.
\newblock {Low background X-ray detection with Micromegas for axion research}.
\newblock {\em Journal of Instrumentation}, 9(01):P01001, 2014.
\newblock \href {https://doi.org/10.1088/1748-0221/9/01/P01001}
  {\path{doi:10.1088/1748-0221/9/01/P01001}}.

\bibitem{Aznar_2015}
F.~Aznar, J.~Castel, F.E. Christensen, T.~Dafni, T.A. Decker, E.~Ferrer-Ribas,
  J.A. Garcia, I.~Giomataris, J.G. Garza, C.J. Hailey, R.M. Hill, F.J. Iguaz,
  I.G. Irastorza, A.C. Jakobsen, G.~Luzon, H.~Mirallas, T.~Papaevangelou, M.J.
  Pivovaroff, J.~Ruz, T.~Vafeiadis, and J.K. Vogel.
\newblock A micromegas-based low-background x-ray detector coupled to a
  slumped-glass telescope for axion research.
\newblock {\em Journal of Cosmology and Astroparticle Physics}, 2015(12):008,
  2015.
\newblock \href {https://doi.org/10.1088/1475-7516/2015/12/008}
  {\path{doi:10.1088/1475-7516/2015/12/008}}.

\bibitem{Garza_2015}
J.~G. Garza, S.~Aune, F.~Aznar, D.~Calvet, J.~F. Castel, F.~E. Christensen,
  T.~Dafni, M.~Davenport, T.~Decker, E.~Ferrer-Ribas, J.~Galán, J.~A. García,
  I.~Giomataris, R~M Hill, F.~J. Iguaz, I.~G. Irastorza, A.~C. Jakobsen,
  D.~Jourde, H.~Mirallas, I.~Ortega, T.~Papaevangelou, M.~J. Pivovaroff,
  J.~Ruz, A.~Tomás, T.~Vafeiadis, and J.~K. Vogel.
\newblock {Low Background Micromegas in CAST}.
\newblock {\em Journal of Physics: Conference Series}, 650(1):012008, 2015.
\newblock \href {https://doi.org/10.1088/1742-6596/650/1/012008}
  {\path{doi:10.1088/1742-6596/650/1/012008}}.

\bibitem{AGET}
S.~Anvar, P.~Baron, B.~Blank, J.~Chavas, E.~Delagnes, F.~Druillole,
  P.~Hellmuth, L.~Nalpas, J.L. Pedroza, J.~Pibernat, E.~Pollacco, A.~Rebii, and
  N.~Usher.
\newblock {AGET, the GET front-end ASIC, for the readout of the Time Projection
  Chambers used in nuclear physic experiments}.
\newblock In {\em 2011 IEEE Nuclear Science Symposium Conference Record}, pages
  745--749, 2011.
\newblock \href {https://doi.org/10.1109/NSSMIC.2011.6154095}
  {\path{doi:10.1109/NSSMIC.2011.6154095}}.

\bibitem{Castel2019}
J.~Castel, S.~Cebri{\'{a}}n, I.~Coarasa, T.~Dafni, J.~Gal{\'{a}}n, F.~J. Iguaz,
  I.~G. Irastorza, G.~Luz{\'{o}}n, H.~Mirallas, A.~{Ortiz de Sol{\'{o}}rzano},
  and E.~Ruiz-Ch{\'{o}}liz.
\newblock {Background assessment for the TREX dark matter experiment}.
\newblock {\em The European Physical Journal C 2019 79:9}, 79(9):1--19, 2019.
\newblock \href {https://doi.org/10.1140/EPJC/S10052-019-7282-6}
  {\path{doi:10.1140/EPJC/S10052-019-7282-6}}.

\bibitem{REST2022}
K.Altenmüller, S.~Cebrián, T.~Dafni, D.~Díez-Ibáñez, J.~Galán,
  J.~Galindo, J.A. García, I.~G. Irastorza, G.~Luzón, C.Margalejo,
  H.~Mirallas, L.~Obis, O.~Pérez, K.~Han, K.~Ni, Y.~Bedfer, B.~Biasuzzi,
  E.~Ferrer-Ribas, D.~Neyret, T.~Papaevangelou, C.~Cogollos, and E.~Picatoste.
\newblock {REST-for-Physics, a ROOT-based framework for event oriented data
  analysis and combined Monte Carlo response}.
\newblock {\em Computer Physics Communications}, 273:108281, 2022.
\newblock \href {https://doi.org/10.1016/j.cpc.2021.108281}
  {\path{doi:10.1016/j.cpc.2021.108281}}.

\bibitem{GIOMATARIS2006405}
I.~Giomataris, R.~{De Oliveira}, S.~Andriamonje, S.~Aune, G.~Charpak, P.~Colas,
  G.~Fanourakis, E.~Ferrer, A.~Giganon, Ph. Rebourgeard, and P.~Salin.
\newblock Micromegas in a bulk.
\newblock {\em Nuclear Instruments and Methods in Physics Research Section A:
  Accelerators, Spectrometers, Detectors and Associated Equipment},
  560(2):405--408, 2006.
\newblock \href {https://doi.org/10.1016/j.nima.2005.12.222}
  {\path{doi:10.1016/j.nima.2005.12.222}}.

\bibitem{Microbulk2010}
S.~Andriamonje, D.~Attie, E.~Berthoumieux, M.~Calviani, P.~Colas, T.~Dafni,
  G.~Fanourakis, E.~Ferrer-Ribas, J.~Galan, T.~Geralis, A.~Giganon,
  I.~Giomataris, A.~Gris, C.~Guerrero Sanchez, F.~Gunsing, F.~J. Iguaz,
  I.~Irastorza, R.~De Oliveira, T.~Papaevangelou, J.~Ruz, I.~Savvidis,
  A.~Teixera, and A.~Tomás.
\newblock Development and performance of microbulk micromegas detectors.
\newblock {\em Journal of Instrumentation}, 5(02):P02001, 2010.
\newblock \href {https://doi.org/10.1088/1748-0221/5/02/P02001}
  {\path{doi:10.1088/1748-0221/5/02/P02001}}.

\bibitem{Cebrian:2010ta}
S.~Cebrián, T.~Dafni, E.~Ferrer-Ribas, J.~Galán, I.~Giomataris, H.~Gómez,
  F.J. Iguaz, I.G. Irastorza, G.~Luzón, R.~{de Oliveira}, A.~Rodríguez,
  L.~Seguí, A.~Tomás, and J.A. Villar.
\newblock {Radiopurity of Micromegas readout planes}.
\newblock {\em Astropart. Phys.}, 34:354--359, 2011.
\newblock \href {https://doi.org/10.1016/j.astropartphys.2010.09.003}
  {\path{doi:10.1016/j.astropartphys.2010.09.003}}.

\bibitem{BiPo_detector}
A.S. Barabash, A.~Basharina-Freshville, E.~Birdsall, S.~Blondel, S.~Blot,
  M.~Bongrand, D.~Boursette, V.~Brudanin, J.~Busto, A.J. Caffrey, S.~Calvez,
  M.~Cascella, S.~Cebrián, C.~Cerna, J.P Cesar, E.~Chauveau, A.~Chopra,
  T.~DafnÃ­, S.~De Capua, D.~Duchesneau, D.~Durand, V.~Egorov, G.~Eurin, J.J.
  Evans, L.~Fajt, D.~Filosofov, R.~Flack, X.~Garrido, H.~Gómez, B.~Guillon,
  P.~Guzowski, K.~Holý, R.~Hodák, A.~Huber, C.~Hugon, F.J. Iguaz, I.G.
  Irastorza, A.~Jeremie, S.~Jullian, M.~Kauer, A.~Klimenko, O.~Kochetov, S.I.
  Konovalov, V.~Kovalenko, K.~Lang, Y.~Lemière, T.~Le Noblet, Z.~Liptak, X.R.
  Liu, P.~Loaiza, G.~Lutter, G.~Luzón, M.~Macko, F.~Mamedov, C.~Marquet,
  F.~Mauger, B.~Morgan, J.~Mott, I.~Nemchenok, M.~Nomachi, F.~Nova, H.~Ohsumi,
  G.~Oliviéro, A.~Ortiz de~Solorzano, R.B. Pahlka, J.~Pater, F.~Perrot,
  F.~Piquemal, P.~Povinec, P.~Přidal, Y.A. Ramachers, A.~Remoto, B.~Richards,
  C.L. Riddle, E.~Rukhadze, R.~Saakyan, R.~Salazar, X.~Sarazin, Yu. Shitov,
  L.~Simard, F.~Šimkovic~nd A.~Smetana, K.~Smolek, A.~Smolnikov,
  S.~Söldner-Rembold, B.~Soulé, I.~Štek, J.~Thomas nd~V.~Timkin, S.~Torre,
  Vl.I. Tretyak, V.I. Tretyak, V.I. Umatov, C.~Vilela nd~V.~Vorobel, D.~Waters,
  and A.~Žukauskas.
\newblock The {B}i{P}o-3 detector for the measurement of ultra low natural
  radioactivities of thin materials.
\newblock {\em JINST}, 12:P06002, 2017.
\newblock \href {https://doi.org/10.1088/1748-0221/12/06/P06002}
  {\path{doi:10.1088/1748-0221/12/06/P06002}}.

\bibitem{Aznar_2013}
F~Aznar, J~Castel, S~Cebrián, T~Dafni, A~Diago, J~A García, J~G Garza,
  H~Gómez, D~González-Díaz, D~C Herrera, F~J Iguaz, I~G Irastorza, G~Luzón,
  H~Mirallas, M~A Oliván, A~Ortiz de, Solorzano, P~Pons, A~Rodríguez, E~Ruiz,
  L~Seguí­, A~Tomás, and J~A Villar.
\newblock Assessment of material radiopurity for rare event experiments using
  micromegas.
\newblock {\em Journal of Instrumentation}, 8(11):C11012, 2013.
\newblock \href {https://doi.org/10.1088/1748-0221/8/11/C11012}
  {\path{doi:10.1088/1748-0221/8/11/C11012}}.

\bibitem{tesis_hector}
Hector Mirallas.
\newblock {\em {Development of Large-Scale Micromegas Planes for Rare Event
  Searches}}.
\newblock PhD thesis, Universidad de Zaragoza, to be submitted, 2024.

\bibitem{xenon_rn_distillation}
M.~Murra, Denny Schulte, C.~Huhmann, and C.~Weinheimer.
\newblock Design, construction and commissioning of a high-flow radon removal
  system for xenonnt.
\newblock {\em The European Physical Journal C}, 82, 2022.
\newblock \href {https://doi.org/10.1140/epjc/s10052-022-11001-9}
  {\path{doi:10.1140/epjc/s10052-022-11001-9}}.

\bibitem{charcoal_modane}
R.~Hodák, F.~Perrot, V.~Brudanin, J.~Busto, M.~Havelcová, J.~Hůlka,
  S.~Jullian, O.~Kochetov, D.~Lalanne, P.~Loaiza, J.~Macl, F.~Mamedov,
  J.~Mizera, R.~Noel, F.~Piquemal, E.~Rukhadze, P.~Rulík, K.~Smolek,
  B.~Soulé, T~Suchá, I.~Svetlík, I.~Štekl, G~Warot, M~Zampaolo, and
  M~Žaloudková.
\newblock {Characterization and long-term performance of the Radon Trapping
  Facility operating at the Modane Underground Laboratory}.
\newblock {\em Journal of Physics G: Nuclear and Particle Physics},
  46(11):115105, 2019.
\newblock \href {https://doi.org/10.1088/1361-6471/ab368e}
  {\path{doi:10.1088/1361-6471/ab368e}}.

\bibitem{Rn_removal_mol_sieves}
A.C. Ezeribe, W.~Lynch, R.R.~Marcelo Gregorio, J.~Mckeand, A.~Scarff, and
  N.J.C. Spooner.
\newblock {Demonstration of radon removal from SF6 using molecular sieves}.
\newblock {\em Journal of Instrumentation}, 12(09):P09025, 2017.
\newblock \href {https://doi.org/10.1088/1748-0221/12/09/P09025}
  {\path{doi:10.1088/1748-0221/12/09/P09025}}.

\bibitem{charcoal}
K.~Pushkin, C.~Akerlof, D.~Anbajagane, J.~Armstrong, M.~Arthurs, J.~Bringewatt,
  T.~Edberg, C.~Hall, M.~Lei, R.~Raymond, M.~Reh, D.~Saini, A.~Sander,
  J.~Schaefer, D.~Seymour, N.~Swanson, Y.~Wang, and W.~Lorenzon.
\newblock Study of radon reduction in gases for rare event search experiments.
\newblock {\em Nuclear Instruments and Methods in Physics Research Section A:
  Accelerators, Spectrometers, Detectors and Associated Equipment},
  903:267--276, 2018.
\newblock \href {https://doi.org/10.1016/j.nima.2018.06.076}
  {\path{doi:10.1016/j.nima.2018.06.076}}.

\bibitem{low_radioact_mol_sieves}
R.R.~Marcelo Gregorio, N.J.C. Spooner, J.~Berry, A.C. Ezeribe, K.~Miuchi,
  H.~Ogawa, and A.~Scarff.
\newblock Test of low radioactive molecular sieves for radon filtration in sf6
  gas-based rare-event physics experiments.
\newblock {\em Journal of Instrumentation}, 16(06):P06024, 2021.
\newblock \href {https://doi.org/10.1088/1748-0221/16/06/P06024}
  {\path{doi:10.1088/1748-0221/16/06/P06024}}.

\bibitem{filters_ieee}
K.~Altenmüller, J.~F. Castel, S.~Cebrián, T.~Dafní, D.~Díez-Ibáñez,
  J.~Galán, J.~Galindo, J.~A. García, I.~G. Irastorza, I.~Katsioulas,
  P.~Knights, G.~Luzón, I.~Manthos, C.~Margalejo, J.~Matthews,
  K.~Mavrokoridis, H.~Mirallas, T.~Neep, K.~Nikolopoulos, L.~Obis, A.~Ortiz~de
  Solórzano, O.~Pérez, B.~Philippou, and R.~Ward.
\newblock Purification efficiency and radon emanation of gas purifiers used
  with pure and binary gas mixtures for gaseous dark matter detectors.
\newblock In {\em 2021 IEEE Nuclear Science Symposium and Medical Imaging
  Conference (NSS/MIC)}, pages 1--3, 2021.
\newblock \href {https://doi.org/10.1109/NSS/MIC44867.2021.9875870}
  {\path{doi:10.1109/NSS/MIC44867.2021.9875870}}.

\bibitem{surface_cont_ptfe}
S.~Bruenner, D.~Cichon, G.~Eurin, P.~Herrero~G\'omez, F.~J\"org,
  T.~Marrod\'an~Undagoitia, H.~Simgen, and N.~Rupp.
\newblock {Radon daughter removal from PTFE surfaces and its application in
  liquid xenon detectors}.
\newblock {\em Eur. Phys. J. C}, 81(4):343, 2021.
\newblock \href {https://doi.org/10.1140/epjc/s10052-021-09047-2}
  {\path{doi:10.1140/epjc/s10052-021-09047-2}}.

\bibitem{alphacamm}
K.~Altenmüller, J.~F. Castel, S.~Cebrián, Th. Dafni, D.~Díez-Ibáñez,
  J.~Galán, J.~Galindo, J.~A. García, I.~G. Irastorza, G.~Luzón,
  C.~Margalejo, H.~Mirallas, L.~Obis, A.~Ortiz de~Solórzano, and O.~Pérez.
\newblock {AlphaCAMM, a Micromegas-based camera for high-sensitivity screening
  of alpha surface contamination}.
\newblock {\em Journal of Instrumentation}, 17(08):P08035, 2022.
\newblock \href {https://doi.org/10.1088/1748-0221/17/08/P08035}
  {\path{doi:10.1088/1748-0221/17/08/P08035}}.

\bibitem{Tomas_2012}
A.~Tomas, S.~Aune, T.~Dafni, G.~Fanourakis, E.~Ferrer-Ribas, J.~Galán, J.A.
  García, A.~Gardikiotis, T.~Geralis, I.~Giomataris, H.~Gómez, J.G. Garza,
  D.C. Herrera, F.J. Iguaz, I.G. Irastorza, G.~Luzón, T.~Papaevangelou,
  A.~Rodríguez, J.~Ruz, L.~Seguí, T.~Vafeiadis, and S.C. Yildiz.
\newblock {CAST Microbulk Micromegas in the Canfranc Underground Laboratory}.
\newblock {\em Physics Procedia}, 37:478–482, 2012.
\newblock \href {https://doi.org/10.1016/j.phpro.2012.02.399}
  {\path{doi:10.1016/j.phpro.2012.02.399}}.

\bibitem{ERuiz}
E.~Ruíz-Chóliz.
\newblock {\em {Ultra-low background Micromegas X-ray detectors for Axion
  searches in IAXO and BabyIAXO}}.
\newblock PhD thesis, University of Zaragoza, 2019.
\newblock URL: \url{https://zaguan.unizar.es/record/87032}.

\bibitem{CMargalejo}
C.~Margalejo.
\newblock {\it{ Modelo de fondo para IAXO-D0, prototipo del experimento IAXO
  (International Axion Observatory)} }.
\newblock Master's thesis, University of Zaragoza, 2018.
\newblock URL: \url{https://zaguan.unizar.es/record/76133}.

\bibitem{Altenmueller-2024}
K.~Altenmüller, J.~F. Castel, S.~Cebrián, T.~Dafni, D.~Díez-Ibañez,
  A.~Ezquerro, E.~Ferrer-Ribas, J.~Galan, J.~Galindo, J.~A. García,
  A.~Giganon, C.~Goblin, I.~G. Irastorza, C.~Loiseau, G.~Luzón, X.~F. Navick,
  C.~Margalejo, H.~Mirallas, L.~Obis, A.~Ortiz de~Solórzano, T.~Papaevangelou,
  0.~Pérez, A.~Quintana, and J.~Ruz anf J.~K.~Vogel.
\newblock {Background discrimination with a Micromegas detector prototype and
  veto system for BabyIAXO}.
\newblock {\em Frontiers. To be published}, 2024.

\bibitem{ATomas}
A.~Tomás.
\newblock {\em {Development of time projection chambers with micromegas for
  Rare Event Searches}}.
\newblock PhD thesis, University of Zaragoza, 2013.
\newblock URL: \url{https://zaguan.unizar.es/record/12540}.

\bibitem{LiquidO:2019mxd}
A.~Cabrera, J.~dos~Anjos A.~Abusleme, T.~J.~C. Bezerra, M.~Bongrand,
  C.~Bourgeois, D.~Breton, C.~Buck, E.~Chauveau J.~Busto, E.~Calvo, M.~Chen,
  P.~Chimenti, F.~Dal Corso, G.~De Conto, S.~Dusini, G.~Fiorentini, C.~Frigerio
  Martins, A.~Givaudan, B.~Gramlich P.~Govoni, M.~Grassi, Y.~Han, J.~Hartnell,
  C.~Hugon, S.~Jiménez, H.~de~Kerret, A.~Le Nevé, P.~Loaiza, J.~Maalmi,
  F.~Mantovani, L.~Manzanillas, C.~Marquet, J.~Martino, D.~Navas-Nicolás,
  H.~Nunokawa, M.~Obolensky, J.~P. Ochoa-Ricoux, G.~Ortona, C.~Palomares,
  F.~Pessina, A.~Pin andJ. C.~C.~Porter, M.~S. Pravikoff, M.~Roche,
  B.~Roskovec, N.~Roy, C.~Santos, S.~Schoppmann, A.~Serafini, L.~Simard,
  M.~Sisti, L.~Stanco, V.~Strati, J.-S. Stutzmann, F.~Suekane, A.~Verdugo,
  B.~Viaud, C.~Volpe, C.~Vrignon, S.~Wagner, and F.~Yermia.
\newblock {Neutrino Physics with an Opaque Detector}.
\newblock {\em Commun. Phys.}, 4:273, 2021.
\newblock \href {https://doi.org/10.1038/s42005-021-00763-5}
  {\path{doi:10.1038/s42005-021-00763-5}}.

\bibitem{Barth_2013}
K.~Barth, A.~Belov, B.~Beltran, H.~Bräuninger, J.M. Carmona, J.I. Collar,
  T.~Dafni, M.~Davenport, L.~Di Lella, C.~Eleftheriadis, J.~Englhauser,
  G.~Fanourakis, E.~Ferrer-Ribas, H.~Fischer, J.~Franz, P.~Friedrich,
  J.~Galán, J.A. García, T.~Geralis, I.~Giomataris, S.~Gninenko, H.~Gómez,
  M.D. Hasinoff, F.H. Heinsius, D.H.H. Hoffmann, I.G. Irastorza, J.~Jacoby,
  K.~Jakovičić, D.~Kang, K.~Königsmann, R.~Kotthaus, K.~Kousouris,
  M.~Krčmar, M.~Kuster, B.~Lakić, A.~Liolios, A.~Ljubičić, G.~Lutz,
  G.~Luzón, D.W. Miller, T.~Papaevangelou, M.J. Pivovaroff, G.~Raffelt,
  J.~Redondo, H.~Riege, A.~Rodríguez, J.~Ruz, I.~Savvidis, Y.~Semertzidis,
  L.~Stewart, K.~Van Bibber, J.D. Vieira, J.A. Villar, J.K. Vogel,
  L.~Walckiers, and K.~Zioutas.
\newblock Cast constraints on the axion-electron coupling.
\newblock {\em Journal of Cosmology and Astroparticle Physics}, 2013(05):010,
  2013.
\newblock \href {https://doi.org/10.1088/1475-7516/2013/05/010}
  {\path{doi:10.1088/1475-7516/2013/05/010}}.

\bibitem{KANE2002}
S.~{Kane}, J.~{May}, J.~{Miyamoto}, I.~{Shipsey}, S.~{Andriamonje},
  A.~{Delbart}, J.~{Derre}, I.~{Giomataris}, and F.~{Jeanneau}.
\newblock {{A study of Micromegas with Preamplification with a Single GEM}}.
\newblock In {\em Advanced Technology - Particle Physics}, pages 694--703,
  2002.
\newblock \href {https://doi.org/10.1142/9789812776464_0098}
  {\path{doi:10.1142/9789812776464_0098}}.

\bibitem{THERS2001133}
D~Thers, Ph~Abbon, J~Ball, Y~Bedfer, C~Bernet, C~Carasco, E~Delagnes, D~Durand,
  J.-C Faivre, H~Fonvieille, A~Giganon, F~Kunne, J.-M.Le Goff, F~Lehar,
  A~Magnon, D~Neyret, E~Pasquetto, H~Pereira, S~Platchkov, E~Poisson, and
  Ph~Rebourgeard.
\newblock Micromegas as a large microstrip detector for the compass experiment.
\newblock {\em Nuclear Instruments and Methods in Physics Research Section A:
  Accelerators, Spectrometers, Detectors and Associated Equipment},
  469(2):133--146, 2001.
\newblock \href {https://doi.org/10.1016/S0168-9002(01)00769-0}
  {\path{doi:10.1016/S0168-9002(01)00769-0}}.

\bibitem{BOUCHEZ2007425}
J.~Bouchez, D.R. Burke, Ch. Cavata, P.~Colas, X.~{De La Broise}, A.~Delbart,
  A.~Giganon, I.~Giomataris, P.~Graffin, J.-Ph. Mols, F.~Pierre, J.-L. Ritou,
  A.~Sarrat, E.~Virique, M.~Zito, E.~Radicioni, R.~{De Oliveira}, J.~Dumarchez,
  N.~Abgrall, P.~Bene, A.~Blondel, A.~Cervera, D.~Ferrere, F.~Maschiocchi,
  E.~Perrin, J.-P. Richeux, R.~Schroeter, G.~Jover, T.~Lux, A.Y. Rodriguez, and
  F.~Sanchez.
\newblock Bulk micromegas detectors for large tpc applications.
\newblock {\em Nuclear Instruments and Methods in Physics Research Section A:
  Accelerators, Spectrometers, Detectors and Associated Equipment},
  574(3):425--432, 2007.
\newblock \href {https://doi.org/10.1016/j.nima.2007.02.074}
  {\path{doi:10.1016/j.nima.2007.02.074}}.

\bibitem{Atlas2022}
J.~Allard, N.~Andari, M.~Anfreville, D.~Attié, E.~Aubernon, S.~Aune,
  H.~Bachacou, F.~Balli, F.~Bauer, J.~Beltramelli, J.~Bennet, T.~Benoit,
  H.~Bervas, T.~Bey, S.~Bouaziz, M.~Boyer, G.~Cara, T.~Chaleil,
  T.~Chevalérias, X.~Coppollani, J.~Costa, G.~Decock, F.~Deliot, D.~Denysiuk,
  D.~Desforge, G.~Disset, G.A. Durand, R.~Durand, J.~Elman, E.~Ferrer-Ribas,
  M.~Fontaine, A.~Formica, J.~Galán, W.~Gamache, A.~Giganon, J.~Giraud, P.F.
  Giraud, G.~Glonti, C.~Goblin, P.~Graffin, J.C. Guillard, S.~Hassani,
  S.~Hervé, S.~Javello, F.~Jeanneau, D.~Jourde, S.~Jurie, M.~Kebbiri,
  T.~Kawamoto, C.~Lampoudis, J.F. Laporte, D.~Leboeuf, M.~Lefèvre, M.~Lohan,
  C.~Loiseau, P.~Magnier, I.~Mandjavidze, J.~Manjarrés, P.~Mas, M.~Mur,
  R.~Nikolaidou, A.~Peyaud, D.~Pierrepont, Y.~Piret, P.~Ponsot, G.~Prono,
  M.~Riallot, F.~Rossi, P.~Schune, T.~Vacher, M.~Vandenbroucke, A.~Vigier,
  C.~Vuillemin, M.~Usseglio, and Z.~Wu.
\newblock The large inner micromegas modules for the atlas muon spectrometer
  upgrade: Construction, quality control and characterization.
\newblock {\em Nuclear Instruments and Methods in Physics Research Section A:
  Accelerators, Spectrometers, Detectors and Associated Equipment},
  1026:166143, 2022.
\newblock \href {https://doi.org/10.1016/j.nima.2021.166143}
  {\path{doi:10.1016/j.nima.2021.166143}}.

\bibitem{Lin_2018}
H.~Lin, D.~Calvet, L.~Chen, X.~Chen, T.~Dafni, C.~Fu, J.~Galan, K.~Han, S.~Hu,
  Y.~Huo, I.G. Irastorza, X.~Ji, X.~Li, X.~Li, J.~Liu, H.~Mirallas, D.~Neyret,
  K.~Ni, H.~Qiao, X.~Ren, Sh. Wang, Si. Wang, Y.~Yang, Y.~Yuan, T.~Zhang, and
  L.~Zhao.
\newblock {Design and commissioning of a 600 L Time Projection Chamber with
  Microbulk Micromegas}.
\newblock {\em Journal of Instrumentation}, 13(06):P06012, 2018.
\newblock \href {https://doi.org/10.1088/1748-0221/13/06/P06012}
  {\path{doi:10.1088/1748-0221/13/06/P06012}}.

\bibitem{Zhang2023}
W.~Zhang, H.~Lin, Y.~Liu, K.~Han, K.~Ni, S.~Wang, W.~Zhai, and the
  PandaX-III~Collaboration.
\newblock {Status and prospects of the PandaX-III experiment}.
\newblock {\em Journal of Instrumentation}, 18(12):C12001, 2023.
\newblock \href {https://doi.org/10.1088/1748-0221/18/12/C12001}
  {\path{doi:10.1088/1748-0221/18/12/C12001}}.

\bibitem{GERALIS:2015bO}
T.~Geralis.
\newblock {A real x-y microbulk Micromegas with segmented mesh}.
\newblock In {\em Proceedings of Technology and Instrumentation in Particle
  Physics 2014 {\textemdash} PoS(TIPP2014)}, volume 213, page 055, 2015.
\newblock \href {https://doi.org/10.22323/1.213.0055}
  {\path{doi:10.22323/1.213.0055}}.

\bibitem{DIAKAKI201846}
M.~Diakaki, E.~Berthoumieux, T.~Papaevangelou, F.~Gunsing, G.~Tsiledakis,
  E.~Dupont, S.~Anvar, L.~Audouin, F.~Aznar, F.~Belloni, E.~Ferrer-Ribas,
  T.~Dafni, D.~Desforge, T.~Geralis, Y.~Giomataris, J.~Heyse, F.J. Iguaz,
  D.~Jourde, M.~Kebbiri, C.~Paradela, P.~Sizun, P.~Schillebeeckx,
  L.~Tassan-Got, and E.~Virique.
\newblock Development of a novel segmented mesh micromegas detector for neutron
  beam profiling.
\newblock {\em Nuclear Instruments and Methods in Physics Research Section A:
  Accelerators, Spectrometers, Detectors and Associated Equipment}, 903:46--55,
  2018.
\newblock \href {https://doi.org/10.1016/j.nima.2018.06.019}
  {\path{doi:10.1016/j.nima.2018.06.019}}.

\bibitem{ALEXOPOULOS2011110}
T.~Alexopoulos, J.~Burnens, R.~{de Oliveira}, G.~Glonti, O.~Pizzirusso,
  V.~Polychronakos, G.~Sekhniaidze, G.~Tsipolitis, and J.~Wotschack.
\newblock A spark-resistant bulk-micromegas chamber for high-rate applications.
\newblock {\em Nuclear Instruments and Methods in Physics Research Section A:
  Accelerators, Spectrometers, Detectors and Associated Equipment},
  640(1):110--118, 2011.
\newblock \href {https://doi.org/10.1016/j.nima.2011.03.025}
  {\path{doi:10.1016/j.nima.2011.03.025}}.

\bibitem{Aznar2020}
F.~Aznar, J.~Castel, S.~Cebri{\'{a}}n, I.~Coarasa, T.~Dafni, J.~Gal{\'{a}}n,
  J.~G. Garza, F.~J. Iguaz, I.~G. Irastorza, G.~Luz{\'{o}}n, H.~Mirallas,
  A.~Ortiz~De Sol{\'{o}}rzano, E.~Ruiz-Ch{\'{o}}liz, and J.~A. Villar.
\newblock {Status of the TREX-DM experiment at the Canfranc Underground
  Laboratory}.
\newblock {\em Journal of Physics: Conference Series}, 1342(1):012091, 2020.
\newblock \href {https://doi.org/10.1088/1742-6596/1342/1/012091}
  {\path{doi:10.1088/1742-6596/1342/1/012091}}.

\end{thebibliography}

}

\end{document}